\documentclass[lettersize,journal]{IEEEtran}
\usepackage{amsmath,amsfonts}
\usepackage{algorithmic}
\usepackage{algorithm}
\usepackage{array}
\usepackage[caption=false,font=normalsize,labelfont=sf,textfont=sf]{subfig}
\usepackage{textcomp}
\usepackage{stfloats}
\usepackage{url}
\usepackage{verbatim}
\usepackage{graphicx}
\usepackage{cite}
\usepackage{booktabs}
\begin{document}

\title{Leveraging Bayesian Optimization for Array Shape \\Self-Calibration in Underwater DoA Estimation}

\author{
Xin~Gui,
Tianang~Li,
Changjia~Wang,
Bowen~Han,\\
Yunchuan~Zhang,~\IEEEmembership{Member,~IEEE},
and~Zhengying~Li

\thanks{
This work was supported in part by the National Nature Science Foundation of China under Grant No. 62471347, the Hubei Provincial Department of Science and Technology under Grant No. 2024BAB022, the Open Fund (2024KF-01) of Hubei Longzhong Laboratory, and Youth Project of Hubei Natural Science Foundation under Grant No. JCZRQNB202600496.
\textit{(Corresponding authors: Yunchuan Zhang)}
}

\thanks{
Xin Gui is with Hubei Longzhong Laboratory, Wuhan University of Technology Xiangyang Demonstration Zone, Xiangyang 441000, China, National Engineering Research Center of Fiber Optic Sensing Technology and Networks, Wuhan University of Technology, Wuhan 430070, China 
(e-mail: guixin@whut.edu.cn).
}

\thanks{
Tianang Li, Changjia Wang, Bowen Han, Yunchuan Zhang, and Zhengying Li are with the School of Information Engineering, 
Wuhan University of Technology, Wuhan 430070, China (email:\{tianangli, wangchangjia, bhan2, yunchuan.zhang, zhyli\}@whut.edu.cn).}

\thanks{
Zhengying Li is also with National Engineering Research Center of Fiber Optic Sensing Technology and Networks, and State Key Laboratory of Advanced Technology for Materials Synthesis and Processing, Wuhan University of Technology, Wuhan, 430070, China.}

}



\maketitle

\begin{abstract}
Flexible sensing arrays are commonly used in underwater acoustic networks, but suppressed by unpredictable geometric deformations. Existing array shape self-calibration methods often estimate individual element positions separately, leading to a high dimensional optimization problem over long arrays. To address this problem, this paper proposes a Bayesian Optimization-assisted Geometry Estimation (BOGE) strategy operating with a hierarchical optimization process and a physics-informed parametric model for array geometry correction. BOGE formulates array shape self-calibration as an optimization problem, where candidate geometries are evaluated by the noise subspace residual. We perform Bayesian optimization to configure the physics-informed parametric model and then refine the selected geometry through numerical optimization. Empirical results show that BOGE achieves lower mean geometric root mean square error (RMSE) than the benchmark methods across a wide range of noise levels. On the public SWellEx-96 dataset, BOGE achieves a geometric RMSE of $0.659$ meters at $166$ Hz. A lake trial further shows that BOGE provides fixed source localization and moving target tracking performance comparable to the comparison methods.
\end{abstract}


\section{Introduction}
\IEEEPARstart{A}{rray} signal processing plays a vital role in the direction of arrival (DoA) estimation for modern Internet of Underwater Things (IoUT) systems, with extensive applications in sonar detection \cite{zhang2022robust}, underwater communications \cite{jahanbakht2021internet,li2025underwater} , and distributed sensing networks \cite{xenaki2025overview}. Array-based estimation methods require the precise knowledge of array manifolds and sensing elements positions. However, flexible arrays, such as towed line arrays \cite{gharib2024modeling,odom2014passive}, acoustic sensitive optical cables \cite{chen2024distributed}, and airborne antennas \cite{santori2007array}, inevitably face the challenges of shape deformations and element position errors in practical deployments. These perturbations are typically induced by dynamic environmental factors including ocean currents, tides, and hardware manufacturing imperfections. Minor deviations in element positions may bring severe DoA errors \cite{sheikh2016near,wang2013tdoa,friedlander2002sensitivity}.

To tackle the above challenges, array calibration techniques including active calibration and self-calibration, are widely introduced in previous studies. Active calibration utilizes cooperative sources with  known directions to estimate array errors \cite{friedlander2002direction,li2006theoretical}. Although highly accurate, this line of work requires precise source placement, which severely limits system flexibility and covertness, thus making it inappropriate for many dynamic IoUT environments. On the other hand, self-calibration can jointly estimate array shape errors and source DoAs without relying on cooperative sources \cite{weiss2002array,zheng2019towed,miao2026shape}. By exploiting spatial features in the received signal, including covariance structure, subspace information, and phase relationships between sensors, self-calibration methods provide a practical alternative in many applications when deploying cooperative acoustic sources is either physically infeasible or operationally undesirable \cite{liu2016sparse,ramamohan2022self}, such as deep ocean exploration and long-term underwater observation \cite{akyildiz2005underwater,heidemann2012underwater,fischer2020operating}.

Existing array shape self-calibration methods are constrained by several critical challenges in practical deployments \cite{mohsan2023recent}, e.g., restrictive assumptions on array geometry, degraded signal features under strong ambient noise, and limited capability to reconstruct complex deformations. Specifically, self-calibration methods may become unreliable when facing low SNR observation and unpredictable deformations caused by tides, currents, and deployment uncertainties in the absence of spatial references. 

Array shape self-calibration also faces a spatial identifiability problem. Theoretical analysis based on the Cramér-Rao Bound (CRB) and Hybrid Cramér-Rao Bound (HCRB) prove that nominally linear arrays can be ambiguous and unidentifiable under small position errors without sufficient prior information \cite{wan2014identifiability,rockah2003array}. To resolve the rotational and scale ambiguities, existing methods have incorporated specific geometric constraints, such as the known distance between a pair of reference elements and known global rotation of the array, to ensure the identifiability of the element positions \cite{santori2009sensor,wang2023self,song2024sensor}. However, reference constraints only remove rigid geometric ambiguities. Estimating the coordinates of elements independently remains underdetermined due to the unobservable deformations from a single source. Optimization algorithms can only return a numerical solution, which still contains unidentifiable deformation modes. These limitations motivate representing the array geometry as a low dimensional parametric model that reduces the unknown degrees of freedom without restricting the array to a single bow model.

In addition to improving identifiability, reducing the degrees of freedom also makes calibration more tractable. Self-calibration typically requires solving nonlinear and high dimensional optimization problems. Recent efforts in sparse Bayesian Learning (SBL) combined with parametric model of the array shape reduce the number of unknown variables \cite{pan2024fast,zheng2020joint}. The study in \cite{yang2025shape} partitions highly deformed arrays into sub-arrays and uses distributed heading measurements to constrain the sensor positions. These methods reduce the number of variables involved in estimation steps. However, when applied to the shape calibration of long arrays, a single bow or low order parametric model may not capture the complex deformations.

In case of highly deformed arrays, large shape mismatches may increase phase differences over $2\pi$, leading to ghost positions in calibration at a single source. To address this problem, prior works in \cite{liu2018wideband,zhen2018array,yang2019joint} use wide-band information or multi-frequency steering models. These methods exploit the invariance of propagation delays over different frequency bins, alongside structural priors like nominal element-spacing and array bending angles, to improve the identifiability of sensor positions. While theoretically effective, they still impose structural assumptions, which require oversimplified, uniform bending models.

In pursuit of enhanced deployment covertness and system flexibility, recent work has explored the use of non-cooperative signals of opportunity \cite{zhang2024array,wu2021enhanced}. By exploiting the spatial coherence and wide-band characteristics of ambient environmental signals such as ship-radiated noise or periodic mechanical sounds, these methods successfully achieve passive array shape calibration tailored for dynamic environments. Despite enhancing system covertness, the inherent low signal-to-noise ratio (SNR) and the highly non-stationary spatial-temporal characteristics of ambient targets inevitably compromise calibration stability, leading to accuracy degradation in harsh acoustic channels.

To achieve reliable array shape estimation without cooperative sources, we introduce a Bayesian optimization-assisted Geometry Estimation (BOGE) method. The main contributions of this paper are summarized as follows:
\begin{itemize}
    \item We propose a novel data-driven self-calibration method that operates over a physics-informed parametric model for array geometry, and a hierarchical optimization framework built upon Bayesian optimization (BO). The proposed framework circumvents conventional optimization bottlenecks by transforming the intractable independent coordinate search into a highly efficient parametric estimation.
    \item To handle the many degrees of freedom in long arrays, calibration proceeds in two stages. A physics-informed parametric model first captures the main deformation, and a low dimensional residual model then refines deviations. This design reduces the search dimension without restricting the array to a single low order shape model.
    \item We evaluate BOGE on the SWellEx-96 dataset and a real-world lake trial using a fiber-optic hydrophone array. At $166$ Hz, BOGE obtains a geometric RMSE of $0.659$ m for the reconstruction on the North horizontal linear array (HLA). In the lake-trial, BOGE localized a fixed source and continuously tracked a moving source over $60$ seconds. Empirical results show that BOGE provides stable and reliable DoA estimation performance for a single source.
\end{itemize}

The rest of the paper is organized as follows. In Section \ref{sec: problem statement}, we present the received data model with shape mismatch and the problem statement of both shape calibration and DoA estimation. Section \ref{sec: array shape calibration method} illustrates the proposed BOGE method. The simulation results of BOGE are discussed in Section \ref{sec: simulation results}. Finally, in Section \ref{sec: real-world validations} we validate the performance of BOGE method in the public dataset SWellEx-96 and a real-world lake trial. 

\section{Problem Statement}\label{sec: problem statement}
In this section, we define the array model with geometry deformations caused by unpredictable environmental factors, including ocean currents, tides, and marine life interference. Then, we introduce the self-calibration framework of interest in the next section, which is referred to as BOGE.

\subsection{Ideal Array Model}\label{sec:array model}

\begin{figure}[!t]
\centering
\includegraphics[width=\columnwidth]{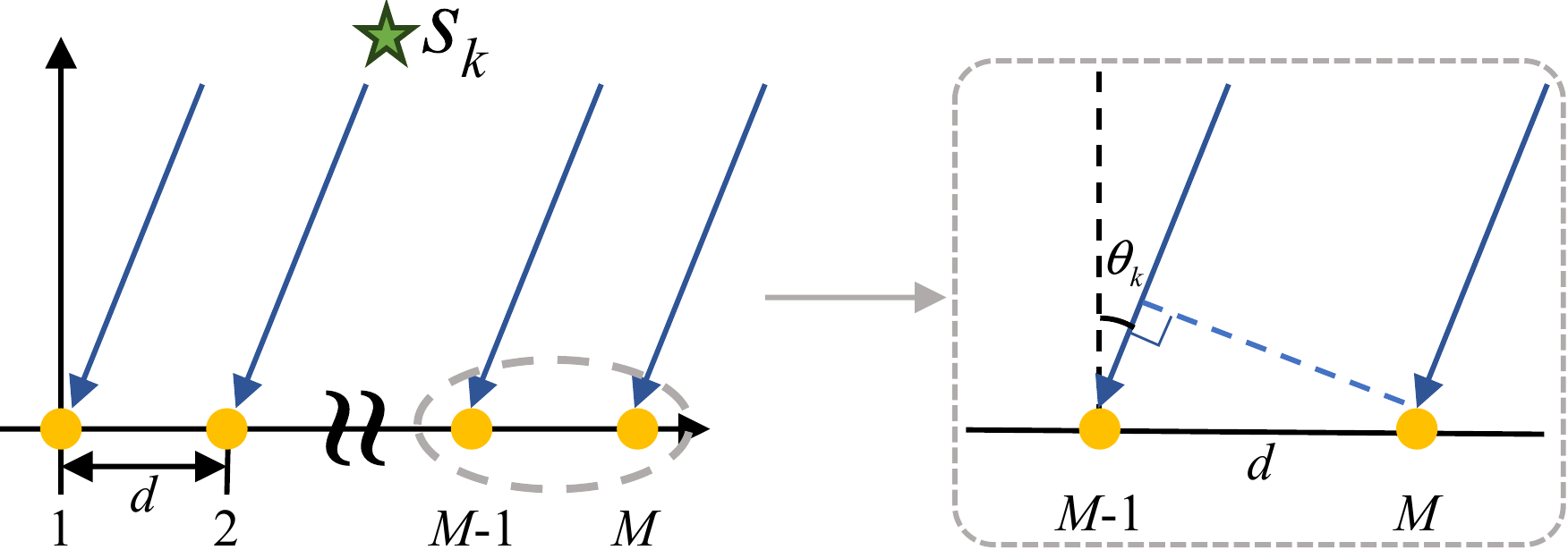}
\caption{Schematic diagram of ULA.}
\label{fig: schematic diagram of ULA}
\vspace{-0.4cm}
\end{figure}

We consider a uniform linear array (ULA) consisting of $M$ sensing elements defined on 2D space. As illustrated in Fig. \ref{fig: schematic diagram of ULA}, the element-spacing is denoted by $d$, and the array aligns with the coordinate axis. Thus, the ideal position of the $m$-th element can be defined as
$
\mathbf{r}_m=[x_m,y_m]^{\sf T}=\big[(m-1)d,0\big]^{\sf T}, \label{eq: sensor 2d location}
$
where the element index $m = 1, 2, \dots, M$.
Assume that $K$ narrow-band acoustic sources are in the far field, influencing the array from directions $\theta_k$. The unit direction vector of the $k$-th incident plane wave can be defined as
$
\mathbf{v}(\theta_k) = [\cos\theta_k , \sin\theta_k]^{\sf T} ,\label{eq: sound wave vector}
$
where the signal index $k = 1, 2, \dots, K$.

Let $s_k(t) = u_k(t)\exp(j2\pi f_ct)$ denote the signal from $k$-th source received at the reference element, where $u_k(t)$ and $f_c$ denote the complex envelope and the carrier frequency of the $k$-th source, respectively. Based on the narrow-band assumption, the variation of the signal envelope over the travel time across the array is negligible, i.e., $u_k(t) \approx u_k(t-\tau_m)$. Consequently, the signal received from element $m$ can be expressed as
\begin{align}
    x_{m,k}(t)& = s_k(t - \tau_m)\nonumber\nonumber\\ 
    &= u_k(t - \tau_m)\exp\big(j2\pi f_c (t - \tau_m)\big) \nonumber\\
    &= u_k(t)\exp({j2\pi f_c t})\exp({-j2\pi f_c \tau_m}) \nonumber\\
    &= s_k(t)\exp({-j2\pi f_c \tau_m})\label{eq: received signal of element m bring in latency}
\end{align}
with $\tau_{m,k}$ representing the time delay between element $m$ and the reference element. 
For a standard ULA, $\tau_{m,k}$ is obtained as
\begin{align}
    \tau_{m,k} = \frac{1}{c}\mathbf{r}_m^{\sf T} \mathbf{v}(\theta_k) = \frac{1}{c}(m-1)d \cos\theta_k,\label{eq: latency expression}
\end{align}
where $c$ represents the speed of sound in water.

By substituting \eqref{eq: latency expression} into \eqref{eq: received signal of element m bring in latency}, the received signal vector for the $k$-th source at time $t$ can be defined as
\begin{align}
    \mathbf{x}_k(t) &= [x_{1,k}(t),\dots, x_{M,k}(t)]^{\sf T}\nonumber\\
    &=\Big[1, \dots, \exp({-j\frac{2\pi}{\lambda}(M-1)d\cos\theta_k})\Big]^{\sf T} s_k(t)\nonumber\\
    &= \mathbf{a}(\theta_k,\mathbf R) s_k(t),\label{eq: received signal vector}
\end{align}
where the steering vector $\mathbf{a}(\theta_k; \mathbf R) \in \mathbb{C}^{M \times 1}$ corresponds to the $k$-th source, and it is dependent on the element position matrix $\mathbf R=[\mathbf r_1,\mathbf r_2,\dots,\mathbf r_M]^{\sf T}$. By incorporating all $K$ incident sources, the received signal matrix can be obtained as
\begin{align}
    \mathbf{x}(t) &= \sum_{k=1}^{K} \mathbf{a}(\theta_k,\mathbf R) s_k(t)
    = \mathbf{A}(\boldsymbol{\theta}, \mathbf R) \mathbf{s}(t),\label{eq: received signal matrix}
\end{align}
where $\mathbf{A}(\boldsymbol{\theta}, \mathbf R) \in \mathbb{C}^{M \times K}$ is the array steering matrix with $\boldsymbol{\theta}$ collecting $[\theta_1,\theta_2,\dots,\theta_K]$, and $\mathbf{s}(t) \in \mathbb{C}^{K \times 1}$ represents the source signal vector.

\subsection{Array Model with Shape Mismatch}\label{sec: array model with shape mismatch}
In most cases, the array is assumed to follow a straight line or a predefined ideal shape. However, the array geometry may change dynamically and cannot be accurately estimated in real-time, resulting in array shape mismatch during array signal processing. Let us define $\tilde{\mathbf{r}}_m$ as the actual position of the $m$-th element
\begin{align}
    \tilde{\mathbf{r}}_m = \begin{bmatrix} \tilde{x}_m \\ \tilde{y}_m \end{bmatrix} = \begin{bmatrix} x_m + \Delta x_m \\ y_m + \Delta y_m \end{bmatrix}=\mathbf{r}_m+\Delta \mathbf{r}_m
    ,\label{eq: element position with array shape mismatch}
\end{align}
where element index $m = 1, 2, \dots, M$.

In this setting, the array is no longer strictly aligned along the coordinate axis. Accordingly, the received signal can be reformulated as
\begin{align}
    \tilde{x}_{m,k}(t) &= s_k(t)\exp\left(
    -j\frac{2\pi}{\lambda}
    \tilde{\mathbf r}_m^{\sf T}\mathbf v(\theta_k)
    \right) \nonumber\\
    &= x_{m,k}(t)\exp\left(
    -j\frac{2\pi}{\lambda}
    \Delta\mathbf r_m^{\sf T}\mathbf v(\theta_k)
    \right).\label{eq: received signal with array shape mismatch}
\end{align}
Each element may encounter a distinct spatial displacement. Intuitively, one can define the corresponding error vector for the $k$-th source as
\begin{align}
    \mathbf e_k = \Big[ &\exp\big({-j\frac{2\pi}{\lambda}(\Delta x_1\cos\theta_k + \Delta y_1\sin\theta_k)}\big),\nonumber\\
    &\exp\big({-j\frac{2\pi}{\lambda}(\Delta x_2\cos\theta_k + \Delta y_2\sin\theta_k)}\big), \nonumber\\
    &\dots, \nonumber\\
    &\exp\big({-j\frac{2\pi}{\lambda}(\Delta x_M\cos\theta_k + \Delta y_M\sin\theta_k)}\big)\Big] ^{\sf T}.\label{eq: phase error vector for k-th source}
\end{align}
Here, the error steering vector is direction-dependent.

The complete error matrix for all $K$ sources can be constructed as $\mathbf E = [\mathbf e_1, \mathbf e_2, \dots, \mathbf e_K]\in \mathbb{C}^{M\times K}$. By incorporating the error matrix $\mathbf E$ into the array signal model \eqref{eq: received signal matrix}, the  received noisy signal can be represented as
\begin{align}
    \tilde{\mathbf{x}}(t) &= \tilde{\mathbf{A}}(\boldsymbol{\theta},\widetilde{\mathbf R})\mathbf{s}(t) + \mathbf{n}(t)\nonumber\\&= [\mathbf{A}(\boldsymbol{\theta},\mathbf R) \odot \mathbf E]\mathbf{s}(t) + \mathbf{n}(t),\label{eq: array signal model with error}
\end{align}
where $\odot$ denotes the Hadamard product, and the random vector $\mathbf{n}(t) \in \mathbb{C}^{M \times 1}$ is the observation noise.

Array shape mismatch introduces additional phase errors into the received signal. If the ideal steering matrix is directly employed for DoA estimation algorithm such as MUSIC, these errors will lead to distortions in the spatial spectrum, thereby degrading the DoA estimation performance.
\subsection{DoA Performance with Array Shape Mismatch}\label{DoA Performance with array shape mismatch}
In this section, we adopt the multiple signal classification (MUSIC) algorithm \cite{schmidt1986multiple} as an example to analyze the impact of array shape mismatch on the performance of DoA estimation. The classical MUSIC algorithm attempts to estimate the DoA by exploiting the orthogonality between the signal subspace $\mathbf{U}_s$ and the noise subspace $\mathbf{U}_n$. The ideal MUSIC spatial spectrum is given by
\begin{align}
    P_{\text{MUSIC}}(\theta) = \frac{1}{\mathbf{a}^{\sf H}(\theta)\mathbf{U}_n\mathbf{U}_n^{\sf H}\mathbf{a}(\theta)},\label{eq: ideal MUSIC spectrum}
\end{align}
where a sharp peak occurs when $\theta = \theta_k$. However, when the array shape is distorted, the received signal is generated by the perturbed steering vector. Here, the distorted steering vector can be expressed as
\begin{align}
    \tilde{\mathbf{a}}(\theta_k) = \mathbf{a}(\theta_k) \odot \mathbf e_k,\label{eq: steering vector with shape mismatch}
\end{align}
where $\mathbf e_k$ denotes the phase error vector induced by array shape mismatch. The observed noise subspace $\tilde{\mathbf{U}}_n$, obtained by the eigen-decomposition of the sample covariance matrix, is orthogonal to the distorted observed steering vector rather than the ideal nominal one. As a result, the original subspace orthogonality is violated, yielding
\begin{align}
\tilde{\mathbf{a}}^{\sf H}(\theta_k)\tilde{\mathbf{U}}_n = [\mathbf{a}(\theta_k) \odot \mathbf e_k]^{\sf H}\tilde{\mathbf{U}}_n = 0.\label{eq: subspace orthogonality}
\end{align}

Since the actual array geometry deviates from the nominal geometry, the ideal steering vector is no longer orthogonal to the observed noise subspace. As a result, the spatial spectrum no longer exhibits a peak centered at the true direction. That is, the spatial spectrum may present severe distortions, including peak location shifts, peak height degradation, and main-lobe broadening.

As recently shown in \cite{miao2026shape}, for sinusoidal distortion on a nominal ULA, a maximum positional error of $3$ m results in a $3^\circ$ DoA estimation bias, while a maximum positional error of $10$ m causes the deviation of the estimation to reach  $10^\circ$. These numerical results indicate that even slight array deformations can lead to severe DoA estimation biases, thereby substantiating the fundamental necessity of array shape calibration. 

Note that the acoustic observations from a single source are invariant under a simultaneous rotation of the array and source direction. Consequently, the reconstructed geometry without reference elements and structural constraints is relative. To define the identifiability problem, we only consider the shape mismatch, with known element ordering and nominal element-spacing. The first, middle, and last elements serve as reference elements, whose coordinates are assumed to be known and noncollinear. These assumptions provide directional and positional references needed to acquire absolute coordinates.

Overall, the goal is to estimate the element coordinates $\widetilde{\mathbf R}$ and the direction $\theta$ given the received noisy observations $\tilde{\mathbf{x}}$ and the coordinates of the reference elements. The source direction and the positions of remaining elements are assumed to be unknown.

\section{Array Shape Calibration Method}\label{sec: array shape calibration method}
As demonstrated in the previous sections, hydrodynamic factors in realistic ocean environments inevitably induce severe degradation in the performance of DoA estimation algorithms. However, existing array shape calibration techniques can be infeasible when applied to large scale arrays. On the one hand, classical active calibration methods require additional calibration sources, which limits its applicability in resource-constrained and dynamic environments. On the other hand, self-calibration methods usually treat the coordinate deviations of all elements as independent variables, leading to an optimization problem with high dimensionality.
\begin{figure*}[t]
\centering
\includegraphics[width=0.75\textwidth]{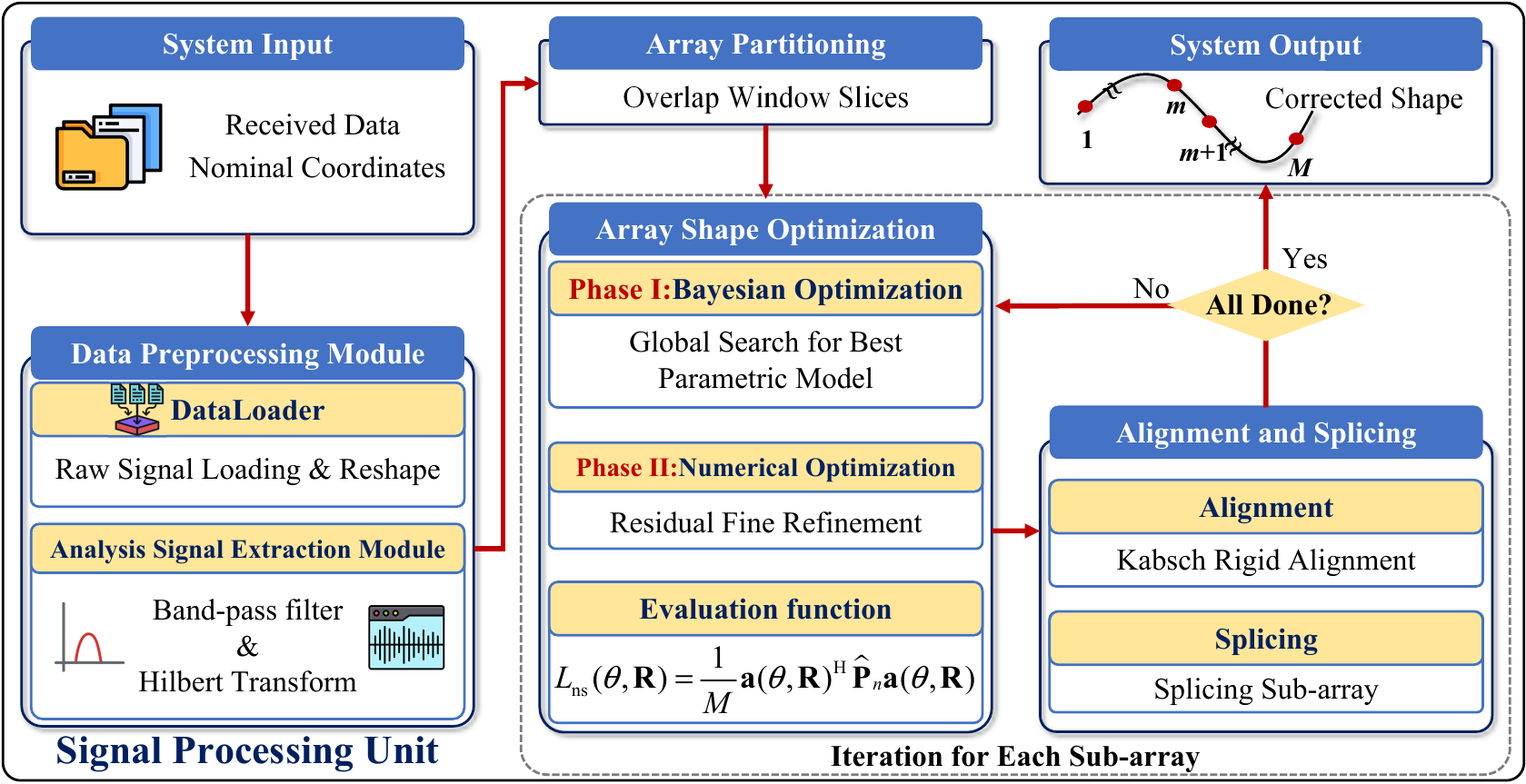}
\caption{BOGE first processes the received data and divide the full array into overlap sub-arrays. A coarse geometry is then estimated by BO in phase I. While phase II refines the residual deformation through numerical optimization. The corrected sub-arrays are aligned and spliced for the reconstruction of long arrays.}
\label{fig: Schematic Diagram for proposed method}
\vspace{-0.4cm}
\end{figure*}

To reduce the dimensionality, the proposed BOGE framework combines a physics-informed parametric model with a coarse-to-fine optimization strategy. As shown in Fig. \ref{fig: Schematic Diagram for proposed method}, the parametric model describes the array geometry using a low dimensional parameters vector, thereby avoiding direct optimization of all element coordinates. BOGE searches feasible curves of the physical model to estimate the coarse shape of global array, and then optimizes the residual coefficients near this coarse estimate to represent deviations. For long arrays, BOGE further adopts overlap sub-array partitioning, followed by alignment and splicing, to reconstruct the full array geometry.

\subsection{Physics-Informed Parameterized Modeling for Array Geometry}\label{sec: Physics-Informed Parameterized Modeling for Array Geometry}
Numerical optimization strategies are often used to calibrate large scale and long aperture arrays shape. For any array consisting of $M$ elements, directly optimizing the spatial coordinates $(x_m,y_m)$ of each individual element yields $2M$ variables. The dimension of parameters to be estimated grows linearly as the number of elements increases, which enlarges the nonconvex search space and computation complexity. To reduce the number of free variables, we introduce the concept of parametric model to describe the array geometry under practical constraints.

Taking the fiber-optic hydrophone array as an example, it is not an arbitrary curve formed by independent discrete points, but rather a continuous physical entity interconnected by optical cables, which possesses inherent tensile strength and bending stiffness. Subject to the effects of complex underwater environments and the structural limitations of the cable, the geometry deformation of the array exhibits the following physical constraints:
\begin{itemize}
    \item \textbf{Spatial Continuity:} The spatial variation along the array geometry is smooth and continuous, precluding any abrupt geometric discontinuities or singular positional perturbations at individual element.
    
    \item \textbf{Distance Boundary Constraints:} The actual inter-element distance, denoted as $l$, is no longer strictly constant and equal to the nominal spacing $d$. Cable slack or stretching allows $l$ to vary within $[d_\text{min}, d_\text{max}]$, where $d_\text{min}<d<d_\text{max}$.
\end{itemize}

Constrained by the physical factors mentioned above, the shape of a large scale fiber-optic hydrophone array can be approximated by a smooth parametric curve. Specifically, a towed array may exhibit periodic sinusoidal vibrations due to hydrodynamic drag, while a stationary array may bend into a circular arc under steady ocean currents. To represent these deformations, let $i=1,\ldots,M-1$ denote the link index along the array. The direction of link $i$ in the local coordinate frame is parameterized as
\begin{align}
    \psi_i(\boldsymbol\eta)
    =\kappa i + A\sin(\omega i+\phi).
    \label{eq: parametric model for array shape}
\end{align}
Here, $\boldsymbol\eta=[A,\omega,\phi,\kappa]^{\sf T}$ is the shape parameter vector. The parameter $\kappa$ describes the arc-like bending, while $A$, $\omega$, and $\phi$ denote the amplitude, spatial frequency, and initial phase of the sinusoidal deformation, respectively.

For each $\boldsymbol\eta$, the local link directions define the local coordinates of the array elements and hence the end-to-end baseline vector $\mathbf d_0(\boldsymbol\eta)$ between the first and last elements. The global rotation $\alpha$ determines the array orientation in the global coordinate frame and is not treated as a BO variable. We obtain it by aligning the direction of $\mathbf d_0(\boldsymbol\eta)$ with that of the known reference vector $\mathbf d_{\text{ref}}$, which gives
\begin{align}
\alpha(\boldsymbol\eta)=
\arg(d_{\text{ref},x}+jd_{\text{ref},y})-\arg(d_{0,x}+j d_{0,y}),\label{eq: analytical global rotation}
\end{align}
where $j=\sqrt{-1}$ and $\arg(\cdot)$ returns the orientation angle of the corresponding two dimensional vector. The direction of link $i$ in the global coordinate frame is then given by
\begin{align}
    \xi_i(\boldsymbol\eta)
    =\alpha(\boldsymbol\eta)+\psi_i(\boldsymbol\eta).
    \label{eq: global link direction}
\end{align}
With $d_i$ denoting the prescribed nominal length of link $i$, the absolute element coordinates can be calculated by
\begin{align}
    \mathbf r_m(\boldsymbol\eta)
    =\mathbf r_1^{\text{ref}}+
    \sum_{i=1}^{m-1}d_i
    \begin{bmatrix}
    \cos\xi_i(\boldsymbol\eta)\\
    \sin\xi_i(\boldsymbol\eta)
    \end{bmatrix},
    \label{eq: coordinate reconstruction from links}
\end{align}
where $m=2,\ldots,M$. The first, middle, and last elements serve as reference elements. After fixing the reference point and the global rotation, feasible parameters should satisfy
\begin{align}
    \|\mathbf d_0(\boldsymbol\eta)\|_2
    &=\|\mathbf d_{\text{ref}}\|_2,\\
    \mathbf r_{m_{\text{c}}}(\boldsymbol\eta)&=\mathbf r_{m_{\text{c}}}^{\text{ref}},
    \label{eq: phase1 reference element constraints}
\end{align}
where $\|\cdot\|_2$ is the Frobenius norm operator; $m_{\text{c}}$ is the index of the middle reference sensor. These equations impose three scalar constraints on four shape parameters. One feasible degree of freedom remains and can be resolved from the acoustic observations.

When $A=0$, the geometry reduces to a circular arc and does not depend on $\omega$ or $\phi$. These two parameters then form an equivalence class. Accordingly, identifiability on this layer concerns the circular arc geometry and its effective curvature, not a unique four dimensional parameter vector.

\subsection{Objective Function Formulation}\label{sec: Objective Function Formulation}
In a practical self-calibration scenario, the true array geometry and the true source DoA are not available for supervision. Therefore, the calibration objective cannot be defined by minimizing the geometric reconstruction error or the DoA estimation error. The objective must be constructed from the received data and the array signal model instead. Based on subspace orthogonality, a matched steering vector is orthogonal to the estimated noise subspace. For a candidate coordinate matrix $\mathbf R$ and a source direction $\theta$, we define the objective function $\mathcal{L}(\cdot)$ as
\begin{align}
    \mathcal{L}(\theta,\mathbf R)
    ={}&\frac{1}{M}
    \mathbf a(\theta,\mathbf R)^{\sf H}
    \widehat{\mathbf P}_n
    \mathbf a(\theta,\mathbf R),
    \label{eq: noise subspace residual}
\end{align}
where $\Theta$ is the admissible direction interval, and $\widehat{\mathbf P}_n=\widehat{\mathbf U}_n\widehat{\mathbf U}_n^{\sf H}$ denotes the estimated noise subspace projector. For each candidate geometry, the direction and objective function is given by
\begin{align}
    \widehat\theta(\mathbf R)
    &=\operatorname*{arg\,min}_{\theta\in\Theta_{\text{g}}}
    \mathcal{L}(\theta,\mathbf R),
    \label{eq: geometry dependent direction estimate}\\
    L(\mathbf R)
    &=\mathcal{L}(\widehat\theta(\mathbf R),\mathbf R)
    =\min_{\theta\in\Theta_{\text{g}}}\mathcal{L}(\theta,\mathbf R),
    \label{eq: profiled noise subspace objective}
\end{align}
where $\Theta_{\text{g}}=\{\theta_1,\ldots,\theta_J\}\subset\Theta$ denote the angular scan grid within the admissible direction interval $\Theta$. Thus, evaluating every candidate geometry returns both its minimum residual $L(\mathbf{R})$ and the corresponding direction $\widehat\theta(\mathbf R)$.

Under the ideal single source model and a correct steering model, the true geometry and direction give $L=0$. With finite snapshots, $\widehat{\mathbf P}_n$ contains estimation error. The true geometry then generally corresponds to a nearby sample minimum.

\subsection{Phase I: BO-Driven Coarse Shape Estimation}\label{sec: BO-Driven Coarse Shape Estimation}
In the coarse estimation phase, we define $L_1(\boldsymbol\eta)=L(\mathbf R(\boldsymbol\eta))$ as the objective function, which is generally nonlinear and nonconvex. The periodic phase terms in the steering vector may produce multiple local minima, making direct local optimization sensitive to initialization. We therefore use BO \cite{zhang2026multi} to obtain a coarse estimate before local refinement phase. However, the reference element constraints restrict the search to a lower dimensional feasible set. These constraints are collected in the vector
\begin{align}
    \mathbf h(\boldsymbol\eta)
    =\begin{bmatrix}
    \|\mathbf d_0(\boldsymbol\eta)\|_2-\|\mathbf d_{\text{ref}}\|_2\\
    \mathbf r_{m_{\rm c}}(\boldsymbol\eta)-\mathbf r_{m_{\rm c}}^{\text{ref}}
    \end{bmatrix}.
    \label{eq: reference element constraint vector}
\end{align}
The corresponding feasible parameter set is defined as
\begin{align}
    \mathcal M
    =\{\boldsymbol\eta:\mathbf h(\boldsymbol\eta)=\mathbf0\}.
    \label{eq: reference element feasible set}
\end{align}
The Jacobian matrix of the constraint vector is defined as
\begin{align}
    \mathbf H(\boldsymbol\eta)
    =\frac{\partial\mathbf h(\boldsymbol\eta)}
    {\partial\boldsymbol\eta}.
    \label{eq: reference element constraint Jacobian}
\end{align}
If $\mathbf H(\boldsymbol\eta)$ has rank three, the implicit function theorem makes $\mathcal M$ locally a one dimensional curve. Rather than searching the original four dimensional parameter box, we construct the feasible set $\mathcal{M}$ numerically and restrict BO to its feasible curves. Multistart root finding on several parameter cross section first identifies witness roots of the reference element equations. Pseudo arclength continuation from these roots then traces a finite set of numerical curves $\mathcal{M}_j$. Each curve is parameterized by the normalized arclength $\zeta_j\in [0,1]$. The witness roots are retained as initial evaluations so that narrow curves are not omitted by the continuation discretization.

BO is performed separately on each curve. The minimum value on curve $j$ is located at
\begin{align}
    \zeta_j^\star
    &=\operatorname*{arg\,min}_{\zeta_j\in[0,1]}
    L_1(\boldsymbol\eta_j(\zeta_j)).
    \label{eq: within curve BO objective}
\end{align}
The curve with the smallest objective value is given by
\begin{align}
    j^\star
    &=\operatorname*{arg\,min}_{j}
    L_1(\boldsymbol\eta_j(\zeta_j^\star)).
    \label{eq: branchwise BO objective}
\end{align}

The one dimensional objective function on curve $j$ is defined as
\begin{align}
    f_j(\zeta)=L_1(\boldsymbol\eta_j(\zeta)).
    \label{eq: branch objective function}
\end{align}
Without loss of generality, we first assume $f_j(\cdot)$ to be a realization of a Gaussian process (GP) with prior
\begin{align}
    f_j(\zeta)\sim\mathcal{GP}(\mu_j(\zeta),k_j(\zeta,\zeta')),
    \label{eq: GP_prior}
\end{align}
where the mean function $\mu_j(\zeta)$ represents the expected objective value, and the kernel function $k_j(\zeta,\zeta')$ describes the correlation between two arclength values.

There are $q$ candidate arclength values evaluated on curve $j$, which are collected in the vector $  \boldsymbol\zeta_{j,q}=(\zeta_{j,1},\ldots,\zeta_{j,q})$. The corresponding objective values are assumed to follow the multivariate Gaussian distribution
\begin{align}
    \begin{bmatrix}
        f_j(\zeta_{j,1})&\cdots&f_j(\zeta_{j,q})
    \end{bmatrix}^{\sf T}
    \sim
    \mathcal N\!\left(
        \boldsymbol\mu_j(\boldsymbol\zeta_{j,q}),
        \mathbf K_j(\boldsymbol\zeta_{j,q})
    \right),
\end{align}
where $\boldsymbol\mu_j(\boldsymbol\zeta_{j,q})$ is a $q\times1$ mean vector and the $q\times q$ covariance matrix $\mathbf K_j(\boldsymbol\zeta_{j,q})$ is given by
\begin{align}
    \mathbf K_j(\boldsymbol\zeta_{j,q})
    =\begin{bmatrix}
        k_j(\zeta_{j,1},\zeta_{j,1})&\cdots&k_j(\zeta_{j,1},\zeta_{j,q})\\
        \vdots&\ddots&\vdots\\
        k_j(\zeta_{j,q},\zeta_{j,1})&\cdots&k_j(\zeta_{j,q},\zeta_{j,q})
    \end{bmatrix}.
    \label{eq: GP covariance matrix}
\end{align}

The choice of the kernel function is pivotal as it encodes the assumed structure of the objective function. In the proposed parametric model, the array geometry varies continuously along each regular curve, and the steering vector is a smooth function of the element positions. Therefore, we adopt the Mat\'ern-$5/2$ kernel \cite{williams2006gaussian}
\begin{align}
    k_j(\zeta,\zeta')
    =\sigma_f^2
    \left(1+\sqrt5\varrho_j+\frac53\varrho_j^2\right)
    \exp(-\sqrt5\varrho_j),
    \label{eq: Matérn-5/2 kernel}
\end{align}
where $\sigma_f^2$ denotes the output variance of the GP surrogate, and $\varrho_j$ represents the Mahalanobis distance, given by
\begin{align}
    \varrho_j=\sqrt{\frac{(\zeta-\zeta')^2}{\chi_j^2}},
    \label{eq: branch scaled distance}
\end{align}
with $\chi_j$ being the length scale for curve $j$.

Building upon the established GP prior, we iteratively update the surrogate model by incorporating new observations. The observations obtained on curve $j$ up to iteration $q$ form the dataset $\mathcal D_{j,q}=\{(\zeta_{j,i},y_{j,i})\}_{i=1}^{q}$.
Each observation is modeled as
\begin{align}
    y_{j,i}=f_j(\zeta_{j,i})+\varepsilon_{j,i},
    \label{eq: GP training data}
\end{align}
with $\varepsilon_{j,i}\sim\mathcal N(0,\sigma_{\text{GP}}^2)$ being the GP observation noise. Given the observation history $\mathcal D_{j,q}$, the posterior distribution at any input $\zeta$ is calculated as
\begin{align}
    p(f_j(\zeta)\mid\mathcal D_{j,q})
    =\mathcal N\!\left(
        \mu_j(\zeta\mid\mathcal D_{j,q}),
        \sigma_j^2(\zeta\mid\mathcal D_{j,q})
    \right),
    \label{eq: GP posterior distribution}
\end{align}
where
\begin{subequations}
\begin{align}
    \mu_j(\zeta\mid\mathcal D_{j,q})
    &=\mu_j(\zeta)
    +\mathbf k_j(\zeta)^{\sf T}
    \widetilde{\mathbf K}_j^{-1}
    \left(
        \mathbf y_{j,q}-\boldsymbol\mu_j(\boldsymbol\zeta_{j,q})
    \right),
    \label{eq: GP posterior mean}\\
    \sigma_j^2(\zeta\mid\mathcal D_{j,q})
    &=k_j(\zeta,\zeta)
    -\mathbf k_j(\zeta)^{\sf T}
    \widetilde{\mathbf K}_j^{-1}
    \mathbf k_j(\zeta).
    \label{eq: GP posterior variance}
\end{align}
\end{subequations}
The covariance vector between the candidate input $\zeta$ and the observed inputs is defined as
\begin{align}
    \mathbf k_j(\zeta)
    =\begin{bmatrix}
        k_j(\zeta_{j,1},\zeta)&\cdots&k_j(\zeta_{j,q},\zeta)
    \end{bmatrix}^{\sf T}.
    \label{eq: GP covariance vector}
\end{align}
The observed objective values are collected in the vector
\begin{align}
    \mathbf y_{j,q}
    =\begin{bmatrix}
        y_{j,1}&\cdots&y_{j,q}
    \end{bmatrix}^{\sf T}.
    \label{eq: GP observation vector}
\end{align}
The covariance matrix including the observation variance is
\begin{align}
    \widetilde{\mathbf K}_j
    =\mathbf K_j(\boldsymbol\zeta_{j,q})+\sigma_{\text{GP}}^2\mathbf I_q.
    \label{eq: GP noisy covariance matrix}
\end{align}

After constructing the GP surrogate, BO determines the next arclength value by optimizing an acquisition function. Since the calibration problem is formulated as a minimization problem, we adopt Expected Improvement (EI) to evaluate the expected magnitude of improvement over the current best observation, which is defined as
\begin{align}
    y_{j,\text{best}}
    =\min_{1\le i\le q}y_{j,i}.
\end{align}
For a candidate arclength value $\zeta^*$, the improvement over the current best observation is
\begin{align}
    I_j(\zeta^*)
    =\max\left(0,y_{j,\text{best}}-f_j(\zeta^*)\right).
\end{align}
Accordingly, the EI acquisition function is defined as
\begin{align}
    \operatorname{EI}_j(\zeta^*)
    =\mathbb E_{f_j(\zeta^*)}
    \left[
        I_j(\zeta^*)\mid\mathcal D_{j,q}
    \right],
    \label{eq: EI definition}
\end{align}
where $f_j(\zeta^*)$ follows the GP posterior defined above. Consequently, the next sampling point is determined by maximizing EI
\begin{align}
    \zeta_{j,q+1}
    =\operatorname*{arg\,max}_{\zeta^*\in[0,1]}
    \operatorname{EI}_j(\zeta^*).
    \label{eq: next point}
\end{align}
The corresponding observation $y_{j,q+1}=f_j(\zeta_{j,q+1})+\varepsilon_{j,q+1}$ is then evaluated, and the dataset is updated as
\begin{align}
    \mathcal D_{j,q+1}
    =\mathcal D_{j,q}
    \cup
    \{(\zeta_{j,q+1},y_{j,q+1})\}.
    \label{eq: dataset update}
\end{align}
Candidate regions with lower predicted objective values are favored. Regions with large posterior uncertainty may also be sampled if they are likely to improve the current best observation.

After BO terminates, we use $\widehat i_j$ to index the smallest observed objective value. The corresponding objective value and arclength are $\widehat y_j=y_{j,\widehat i_j}$ and $\widehat\zeta_j=\zeta_{j,\widehat i_j}$, respectively. The final curve is the one with the smallest observed objective value
\begin{align}
    \widehat j
    =\operatorname*{arg\,min}_{j}
    \widehat y_j.
    \label{eq: selected feasible branch}
\end{align}
The coarse parameter estimate is then reconstructed from the selected curve
\begin{align}
    \widehat{\boldsymbol\eta}_{\text{coarse}}
    =\boldsymbol\eta_{\widehat j}(\widehat\zeta_{\widehat j}).
    \label{eq: bo coarse estimate}
\end{align}
Hereby, we can obtain the corresponding coarse estimates of geometry and source direction as
\begin{align}
    \widehat{\mathbf R}_{\text{coarse}}
    &=\mathbf R(\widehat{\boldsymbol\eta}_{\text{coarse}}),
    \label{eq:phase 1 geometry outputs}
    \\
    \widehat\theta_{\text{coarse}}
    &=\widehat\theta(\widehat{\mathbf R}_{\text{coarse}}).
    \label{eq:phase 1 direction outputs}
\end{align}

The following condition characterizes local identifiability in the coarse estimation phase. After referencing the phases to the first element, the relative phase vector is defined as
\begin{align}
    \mathbf g(\boldsymbol\eta,\theta)
    =-\frac{2\pi}{\lambda}
    \begin{bmatrix}
    [\mathbf r_2(\boldsymbol\eta)-\mathbf r_1(\boldsymbol\eta)]^{\sf T}\mathbf v(\theta)\\
    \vdots\\
    [\mathbf r_M(\boldsymbol\eta)-\mathbf r_1(\boldsymbol\eta)]^{\sf T}\mathbf v(\theta)
    \end{bmatrix}.
    \label{eq: phase1 relative phase vector}
\end{align}
At the true geometry and source direction, let $\mathbf{G}_\eta$ denote the relative phase sensitivity matrix for the shape parameters, and $\mathbf{G}_\theta$ denotes the corresponding sensitivity vector for the source direction, which can be defined as
\begin{align}
    \mathbf G_\eta
    &=\left.
    \frac{\partial\mathbf g(\boldsymbol\eta,\theta)}
    {\partial\boldsymbol\eta^{\sf T}}
    \right|_{(\boldsymbol\eta_\star,\theta_\star)},
    \label{eq: phase1 relative phase sensitivities 1}
    \\
    \mathbf G_\theta
    &=\left.
    \frac{\partial\mathbf g(\boldsymbol\eta,\theta)}
    {\partial\theta}
    \right|_{(\boldsymbol\eta_\star,\theta_\star)},
    \label{eq: phase1 relative phase sensitivities 2}
\end{align}
for some regular true parameter $\boldsymbol\eta_\star$ with amplitute $A_\star>0$, and let the Jacobian matrix of constraint vector $\mathbf H_\star=\mathbf H(\boldsymbol\eta_\star)$ satisfy $\operatorname{rank}[\mathbf H_\star]=3$. Matrices in \eqref{eq: phase1 relative phase sensitivities 1} and \eqref{eq: phase1 relative phase sensitivities 2} quantify the first order changes in the relative phases caused by shape and direction perturbations, respectively. The tangent direction $\boldsymbol\nu_\star$ of the feasible curve satisfies $\mathbf H_\star\boldsymbol\nu_\star=\mathbf0$. If
\begin{align}
    \operatorname{rank}
    \begin{bmatrix}
    \mathbf G_\eta\boldsymbol\nu_\star&\mathbf G_\theta
    \end{bmatrix}=2,
    \label{eq: phase1 local identifiability rank}
\end{align}
the feasible shape perturbation cannot be canceled by a direction perturbation. Thus, $\boldsymbol\eta_\star$ is a strict isolated local minimum of the population objective on $\mathcal M$.

Global uniqueness requires additional conditions. Suppose $\mathcal M$ contains finitely many regular curves, the relative phase map $\mathbf g(\boldsymbol\eta,\theta)$ is injective on each $\mathcal M_j\times\Theta$, and the images of different curve domains do not intersect. Under these conditions, the true geometry is the unique population zero across the feasible set. When $A_\star=0$, the same statement applies to the circular arc geometry class because $\omega$ and $\phi$ remain inactive.

\begin{figure*}[t]
\centering

\includegraphics[width=\textwidth]{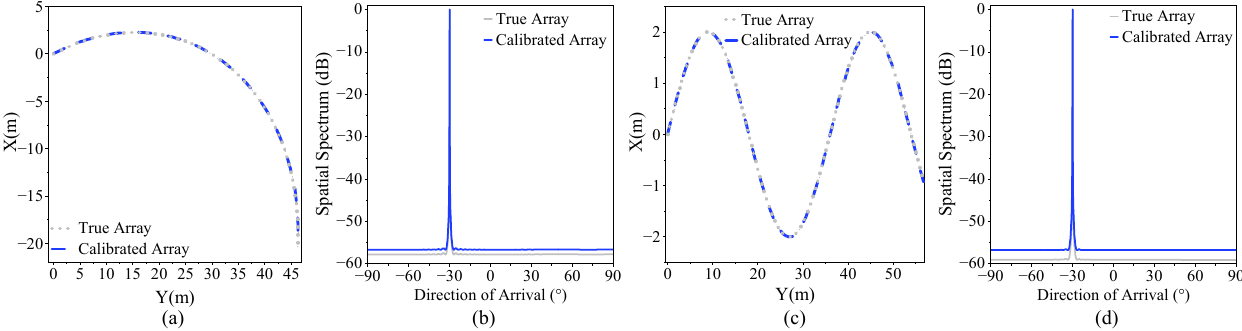}

\caption{Comparison of array geometry reconstruction and MUSIC spatial spectrum under different scenarios: (a) 2D geometry estimation of the circular-arc array. (b) Spatial spectrum comparison for the circular-arc array. (c) 2D geometry estimation of the sinusoidal array. (d) Spatial spectrum comparison for the sinusoidal array.}
\label{fig: simulation results}
\vspace{-0.4cm}
\end{figure*}

\subsection{Phase II: Fine Shape Refinement}\label{sec: BFGS-Based Fine Shape Refinement}
To achieve high precision parameter estimation, the proposed method introduces a numerical optimization fine refinement stage that builds upon the preliminary array geometry established by the BO coarse estimate. Although this estimate captures the array coarse geometry, the four dimensional model may not represent small deviations from it.

Specifically, the fine estimation phase adopts the coarse estimate yielded by the BO output as its initialization, and refines the estimate by optimizing a residual model. This estimate places the numerical refinement in a locally informative region. Within this localized region, gradient-based optimization algorithms enable efficient refinement of the residual coefficients. Furthermore, recognizing that residual coefficients and shape deformation are restricted by physical boundaries, we formulate the fine estimation as a constrained numerical optimization problem. The refined coordinate matrix is written as
\begin{align}
    \mathbf R(\boldsymbol\gamma)
    =\widehat{\mathbf R}_{\text{coarse}}
    +\text{Mat}_{M\times2}
    (\mathbf B\boldsymbol\gamma).
    \label{eq: phase2 residual model}
\end{align}
Here, $\mathbf B\in\mathbb R^{2M\times p}$ is the residual basis; $p$ denotes the number of residual coefficients that are observable at the coarse estimate; $\boldsymbol\gamma\in\mathbb R^p$ collects the residual coefficients. Each column of $\mathbf B\in\mathbb R^{2M\times p}$ represents a deformation mode over the entire array, while each two row blocks maps these coefficients to the two dimensional coordinate correction of each element. The operator $\text{Mat}_{M\times2}(\cdot)$ reshapes a length $2M$ vector into an $M\times2$ matrix.

To construct the residual basis matrix $\mathbf{B}$, we generate smooth candidate corrections using cubic B-splines along the tangent and normal directions of the coarse geometry. The corrections at the three reference elements are set to zero, which preserves the coordinates of the reference elements.

In numerical refinement phase, the coarse estimate of source direction $\widehat\theta_{\text{coarse}}$ obtained by \eqref{eq:phase 1 direction outputs} is fixed. We formulate the fine shape refinement phase as an optimization over the residual coefficients with objective
\begin{align}
    L_2(\boldsymbol\gamma)
    ={}&\frac{1}{M}
    \mathbf a(\widehat\theta_{\text{coarse}},\mathbf R(\boldsymbol\gamma))^{\sf H}
    \widehat{\mathbf P}_n
    \mathbf a(\widehat\theta_{\text{coarse}},\mathbf R(\boldsymbol\gamma)).
    \label{eq: phase2 fixed direction objective}
\end{align}

The admissible set imposes coefficient, displacement, and link length constraints. The root mean square coordinate displacement is defined as
\begin{align}
    \delta_{\text{rms}}(\boldsymbol\gamma)
    =\sqrt{
    \frac{1}{M}\sum_{m=1}^{M}
    \|\boldsymbol\delta_m(\boldsymbol\gamma)\|_2^2
    },
    \label{eq: phase2 displacement rms}
\end{align}
where $\boldsymbol\delta_m(\boldsymbol\gamma)$ is the coordinate correction of element $m$. Let $\overline{\boldsymbol\ell}_i$ and $d_i$ denote the link $i$ of the coarse estimate and its nominal length,  respectively. The matrix $\mathbf D_i\in\mathbb R^{2\times2M}$ maps the stacked coordinate corrections to the change in this link. For $j=1,\ldots,p$ and $i=1,\ldots,M-1$, the feasible set $\mathcal{F}_2$ is
\begin{align}
    \mathcal F_2=\left\{\boldsymbol\gamma:\;
    \begin{array}{l}
    |\gamma_j|\le0.25d_{\rm s},\\
    \delta_{\text{rms}}(\boldsymbol\gamma)\le\rho d_{\text{s}},\\
    0.98d_i\le
    \|\overline{\boldsymbol\ell}_i
    +\mathbf D_i\mathbf B\boldsymbol\gamma\|_2
    \le1.02d_i
    \end{array}
    \right\}.
    \label{eq: phase2 feasible coefficient set}
\end{align}
Here, $d_{\rm s}$ denotes the median nominal spacing for array; $\rho$ is the displacement trust ratio, which is set to $0.08$ in the current implementation. These constraints restrict the refinement to small physically admissible deviations from the structure obtained in Phase I.

\begin{figure}[t]
\centering
\includegraphics[width=0.72\columnwidth]{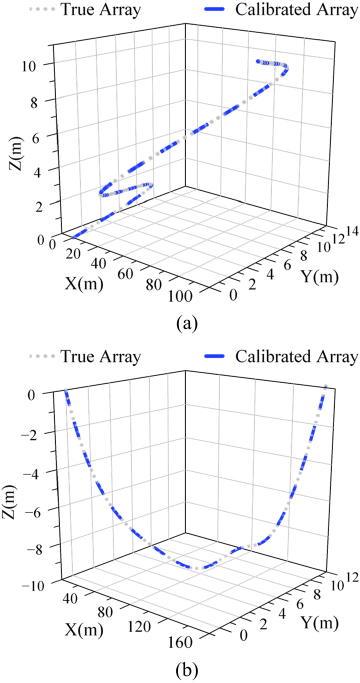}
\caption{Performance on array geometry reconstruction in 3D space: (a) 3D geometry estimation of the sinusoidal array. (b) 3D geometry estimation of the catenary array.}
\label{fig: 3d simulation results}
\vspace{-0.4cm}
\end{figure}

We minimize the objective $L_2(\boldsymbol\gamma)$ in phase II numerically from four initial vectors. The first is $\boldsymbol\gamma_0^{(0)}=\mathbf0$, which corresponds to the coarse estimate of geometry. The other initial points are randomly set in different unit directions within the coefficient space. Each run uses the objective gradient and terminates at a tolerance of $10^{-13}$ or after $500$ iterations. Among the candidates satisfying the constraints within $10^{-8}$, we retain the one with the smallest value of $L_2(\boldsymbol\gamma)$. If none meets this tolerance, the coarse estimate of geometry is retained. In addition, the selected coefficient vector is denoted by $\widehat{\boldsymbol\gamma}_{\text{P2}}$, and the corresponding geometry is $\widehat{\mathbf R}_{\text{P2}}=\mathbf R(\widehat{\boldsymbol\gamma}_{\text{P2}})$.

\begin{figure}[t]
\centering
\includegraphics[width=0.8\columnwidth]{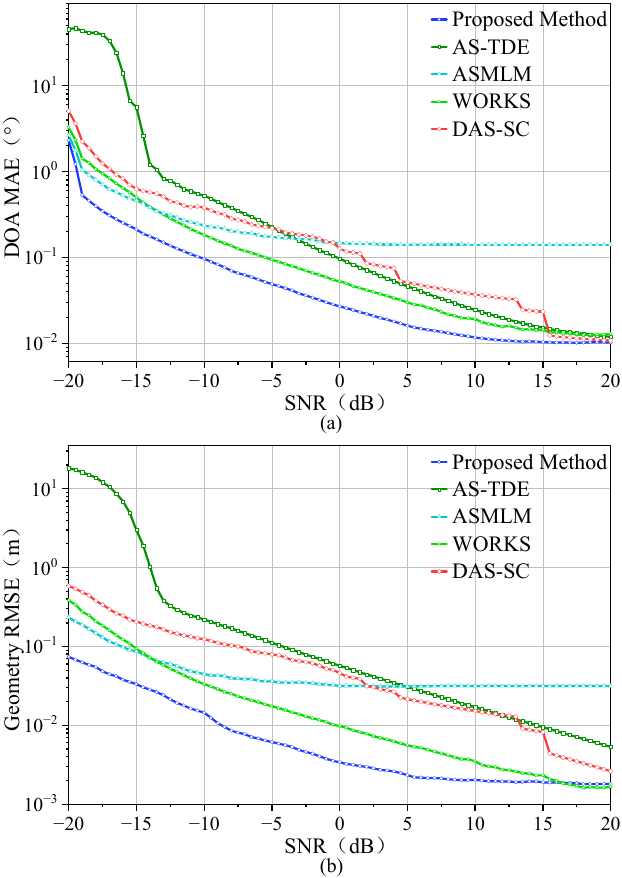}

\caption{Benchmark performance under different SNRs: (a) DoA MAE under different SNRs. (b) Geometric RMSE under different SNRs.}
\label{fig: MC results SNR}
\vspace{-0.4cm}
\end{figure}

For a long array, we apply this complete coarse-to-fine procedure to each overlap sub-array. Then, the shared elements is used to estimate the rigid transformation between adjacent sub-arrays by Kabsch alignment. Coordinates in the overlap are fused to form the full array estimate.

We then examine identifiability of the phase II residual model. Let $\mathbf S\in\mathbb R^{2(M-3)\times2M}$ select the coordinates of the elements other than the three reference elements. For a source direction $\theta$, the corresponding coordinate projection matrix is
\begin{align}
    \mathbf P_\theta
    =\mathbf I_{M-3}\otimes\mathbf v(\theta)^{\sf T},
\end{align}
where $\mathbf I_{M-3}$ is the identity matrix and $\mathbf v(\theta)$ is the direction vector defined in \eqref{eq: sound wave vector}. The projected residual basis is
\begin{align}
    \mathbf C_\theta
    &=\mathbf P_\theta\mathbf S\mathbf B
    \in\mathbb R^{(M-3)\times p}.
    \label{eq: phase2 projected residual matrix}
\end{align}
A single direction provides one scalar projection for each element other than the three reference elements. Thus, for the true source direction $\theta_\star$, identifiability requires
\begin{align}
    p\le M-3,\label{eq: phase2 identifiability dimensian}
    \\
    \operatorname{rank}[\mathbf C_{\theta_\star}]=p.
    \label{eq: phase2 identifiability rank}
\end{align}
Assume that the reference elements are noncollinear, phase ambiguities are absent, and the true residual deformation can be expressed as a linear combination of the columns of $\mathbf B$. The reference elements then determine the unknown source direction. Any other coefficient vector $\boldsymbol\gamma$ that gives zero population residual must satisfy
\begin{align}
    \mathbf C_{\theta_\star}
    (\boldsymbol\gamma-\boldsymbol\gamma_\star)=\mathbf0 ,
\end{align}
where $\boldsymbol\gamma_\star$ denotes the corresponding coefficient vector. If $\mathbf C_{\theta_\star}$ has full column rank, this equation permits only $\boldsymbol\gamma=\boldsymbol\gamma_\star$. The residual coefficients and refined geometry are therefore unique under the stated conditions. If the columns of $\mathbf B$ can only approximate the true residual deformation, Phase 2 recovers a approximation rather than the exact geometry.

\begin{figure}[t]
\centering
\includegraphics[width=\columnwidth]{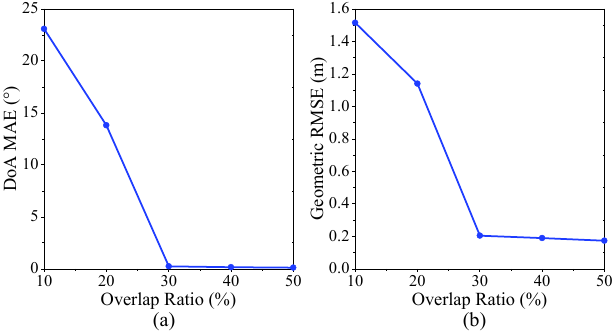}
\caption{Statistical performance of BOGE under different conditions: (a) DoA MAE under different overlap ratios. (b) Geometric RMSE under different overlap ratios.}
\label{fig: MC results overlap}
\vspace{-0.4cm}
\end{figure}

The dimension condition also explains why elementwise refinement is not identifiable from one direction. Fixing three reference elements leaves $2M-6$ coordinate degrees of freedom, whereas one direction provides at most $M-3$ independent constraints. Thus, at least $M-3$ continuous null directions generally remain. Bounds and smoothness penalties may select a numerical solution, but they do not create identifiability from the data.

\section{Simulation Results}\label{sec: simulation results}
In this section, we set up numerical simulations to validate the performance of the proposed BOGE scheme. We first define the comparison methods and their implementations. Then we verify the geometric accuracy of the array shape calibration and the resulting DoA estimation performance in both 2D and 3D spaces. Finally, we compare the five methods across different SNRs and examine the effect of the overlap ratio on BOGE.

\subsection{Benchmark}\label{sec: benchmark}
We select four representative calibration methods to evaluate the performance of BOGE. Recent work in \cite{wang2025adaptive} introduces an adaptive shape estimation method based on marginal likelihood maximization, which we refer to as ASMLM. The opportunity source method in \cite{wu2021enhanced} uses a weighted outlier robust Kalman smoother, abbreviated as WORKS, to estimate array geometry from multiple spectral lines. We denote the auxiliary source calibration method based on time delay estimation in \cite{jun2007method} as AS-TDE. Previous work in \cite{miao2026shape} introduces a self-calibration method for distributed acoustic sensing (DAS) cable, which we refer to as DAS-SC. BOGE uses the implementation developed in this work, whereas the four baselines are independent reproductions based on the published equations. The experiment provides a representative comparison under a common evaluator.

\begin{figure}[t]
\centering
\includegraphics[width=\columnwidth]{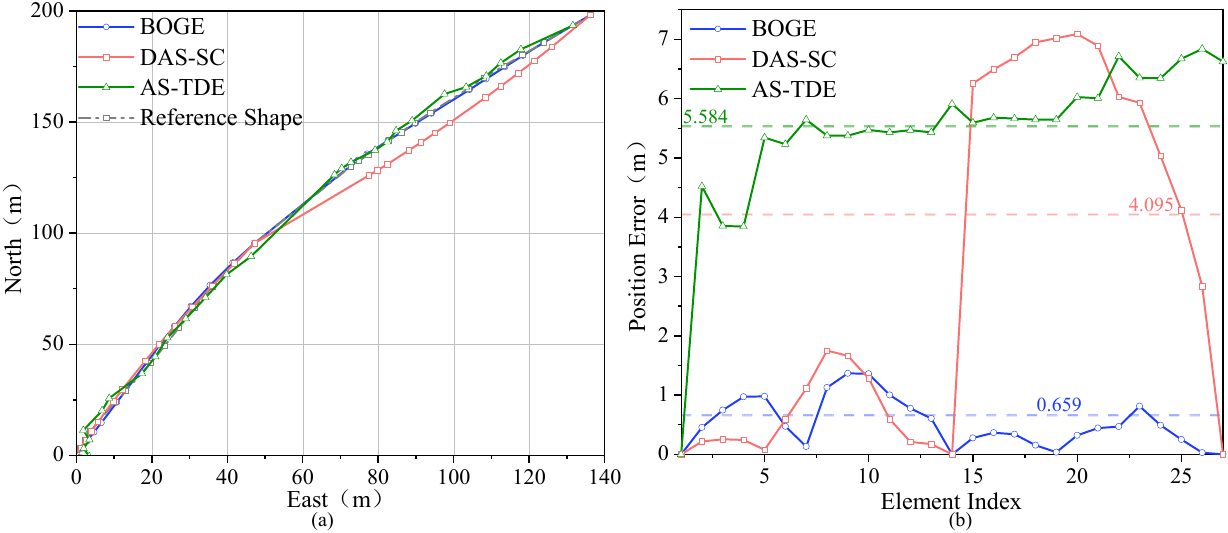}
\caption{Array geometry recovery on the SWellEx-96 dataset: (a) Reference geometry and geometries estimated by BOGE, DAS-SC, and AS-TDE. (b) Position error of each effective element relative to the reference coordinates. }
\label{fig: swellex96 result}\vspace{-0.4cm}
\end{figure}

\subsection{Evaluation of Array Shape Calibration}\label{sec: evaluation of array shape calibration}
To verify the geometric reconstruction capability of the BOGE, a series of numerical simulations are conducted. We first consider a flexible array consisting of $M=40$ elements with a nominal element spacing of $d=1$ meter in 2D space. To emulate the hydrodynamic drag typically experienced in realistic underwater environments, two distinct deformation scenarios are constructed:
\begin{itemize}
    \item The circular-arc shape representative of steady current deflection.

    \item The sinusoidal shape characteristic of wave-induced vibration.
\end{itemize}

Fig. \ref{fig: simulation results} illustrates the calibration performance for the two scenarios. The straight nominal array is omitted from the spatial plots. As shown in Fig. \ref{fig: simulation results}(a) and \ref{fig: simulation results}(c), the calibrated array geometries, denoted by the blue dashed lines, fit the ground truth shapes, represented by the light gray dotted lines. The high degree of geometric fit demonstrates the algorithm's capability to estimate physical deformations with high accuracy. Furthermore, Fig. \ref{fig: simulation results}(b) and \ref{fig: simulation results}(d) demonstrate the corresponding MUSIC spectra, assuming a single incident source from $-30^\circ$. The spatial spectra generated by calibrated arrays can closely match the ideal spectra of the true arrays. Both exhibit sharp and prominent main-lobes at the target direction, effectively eliminating the DoA estimation failure that would occur with the uncalibrated array.

Building upon the 2D verification, the performance of the proposed framework is further evaluated in 3D deformation scenarios. A flexible large scale array consisting of $M=100$ elements with a nominal element-spacing of $d=1$ meter is considered. In this case, the array undergoes simultaneous bending and global axial rolling. Specifically, two representative configurations are constructed: 
\begin{itemize}
    \item A catenary shape emulating that the array is suspended by buoys at both ends.
    \item A tilted sinusoidal geometry representing a configuration where one end is anchored while the array droops and simultaneously exhibits wave-induced sinusoidal vibration.
\end{itemize}

As shown in Fig. \ref{fig: 3d simulation results}, the actual arrays are represented by light gray dotted lines, while calibrated arrays are denoted by blue dashed lines. The calibrated shapes fit the true shapes with high precision. This high degree of geometric alignment is attributed to the efficacy of the parametric model coupled with the overlapped alignment technique, which successfully reduces the estimation error typically associated with long arrays. These 2D and 3D simulations support the effectiveness of the design of BOGE, and motivate the subsequent statistical evaluation.

\begin{figure}[t]
\centering
\includegraphics[width=\columnwidth]{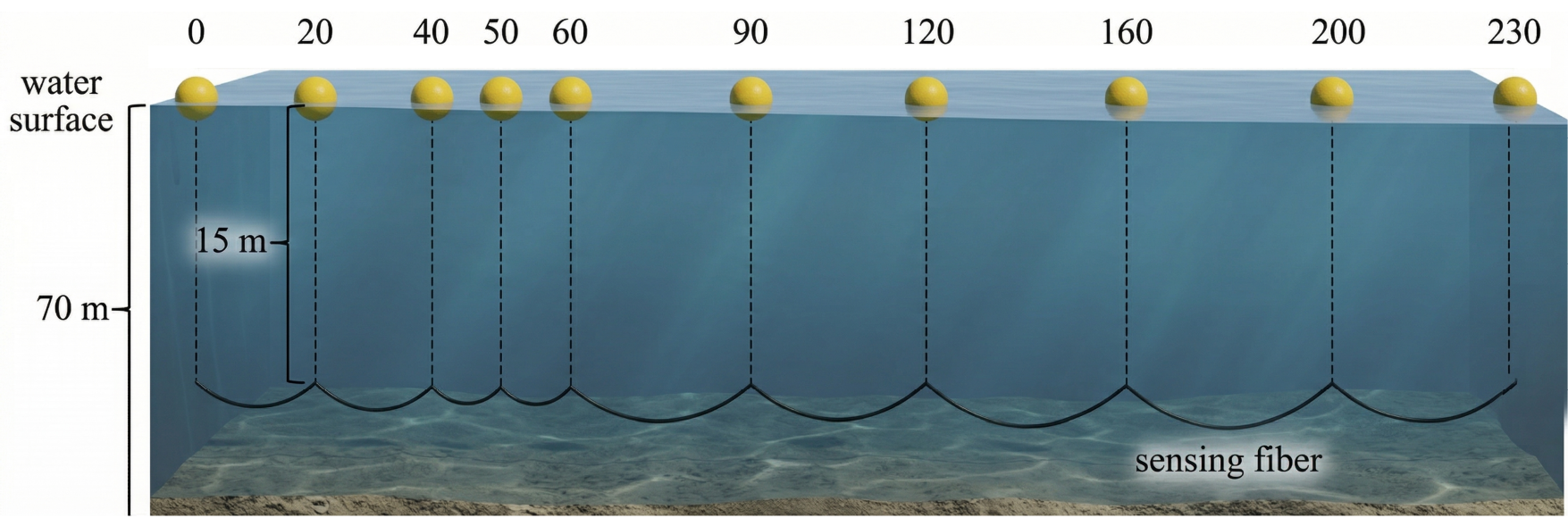}
\caption{Illustration diagram of experimental deployment.}
\label{fig: schematic diagram of array layout}\vspace{-0.4cm}
\end{figure}

\subsection{Statistical Performance of BOGE}\label{sec: statistical performance of proposed method}
While the preceding simulation results illustrate the geometric reconstruction capability of the proposed framework, practical underwater environments are inherently more complicated. To evaluate the statistical robustness and engineering applicability of BOGE, Monte Carlo simulations are conducted in this subsection.

The performance is assessed on an array consisting of $M=20$ elements in 2D space. We first evaluate the algorithm's performance under ambient noise by analyzing the geometric root-mean-square error (RMSE) and the resultant DoA estimation mean absolute error (MAE) under SNRs, ranging from $-20$ dB to $20$ dB in $0.5$ dB increments. For each SNR, $N=200$ Monte Carlo trials are conducted. BOGE and ASMLM share the same narrow-band source in far field. WORKS and AS-TDE retain calibration sources required by their respective formulations. For each experiment, the reported geometric RMSE is the arithmetic mean of the RMSE computed in each trial. The metrics are defined as
\begin{align}
    \overline{\text{RMSE}}_{\text{geo}} &= \frac{1}{N}\sum_{n=1}^{N}\sqrt{\frac{1}{M}\sum_{i=1}^{M}\left\|\widehat{\mathbf r}_{i,n}-\mathbf r_i\right\|^2},\label{eq: geometric RMSE}\\
    \text{MAE}_{\text{DoA}} &= \frac{1}{N} \sum_{n=1}^{N} | \hat{\theta}_n - \theta_{\text{true}} |,\label{eq: DoA estimation MAE}
\end{align}
where $M$ is the total number of elements; the vector $\widehat{\mathbf r}_{i,n}$ and $\mathbf r_i$ denote the estimated coordinates and the true coordinates of the element $i$ in the $n$-th trial, respectively; $\hat{\theta}_n$ represents the estimated DoA in the $n$-th trial, and $\theta_{\text{true}}$ signifies the true direction of the incident source.

We compare the proposed BOGE algorithm with the aforementioned benchmarks in
Fig.\ref{fig: MC results SNR}. All schemes generally achieve more accurate estimation as the SNR increases. While BOGE achieves the lowest mean geometry RMSE over most of the tested range, with respective values of approximately $0.0737$, $0.0145$, and $0.00341$ meters at SNR $-20$, $-10$, and $0$ dB. AS-TDE shows the largest errors in the low SNR region due to the low quality of the calibration signals. WORKS improves as its multitone delay estimates become more reliable. WORKS and BOGE achieve similar mean geometry RMSEs at high SNRs, while BOGE achieves a lower mean DoA MAE. DAS-SC obtains stepped reductions in estimation error as the source SNR increases, due to its sensitivity to source quality. Overall, BOGE maintains low mean geometry and DoA errors across a broad SNR range. 

\begin{figure}[t]
\centering
\includegraphics[width=\columnwidth]{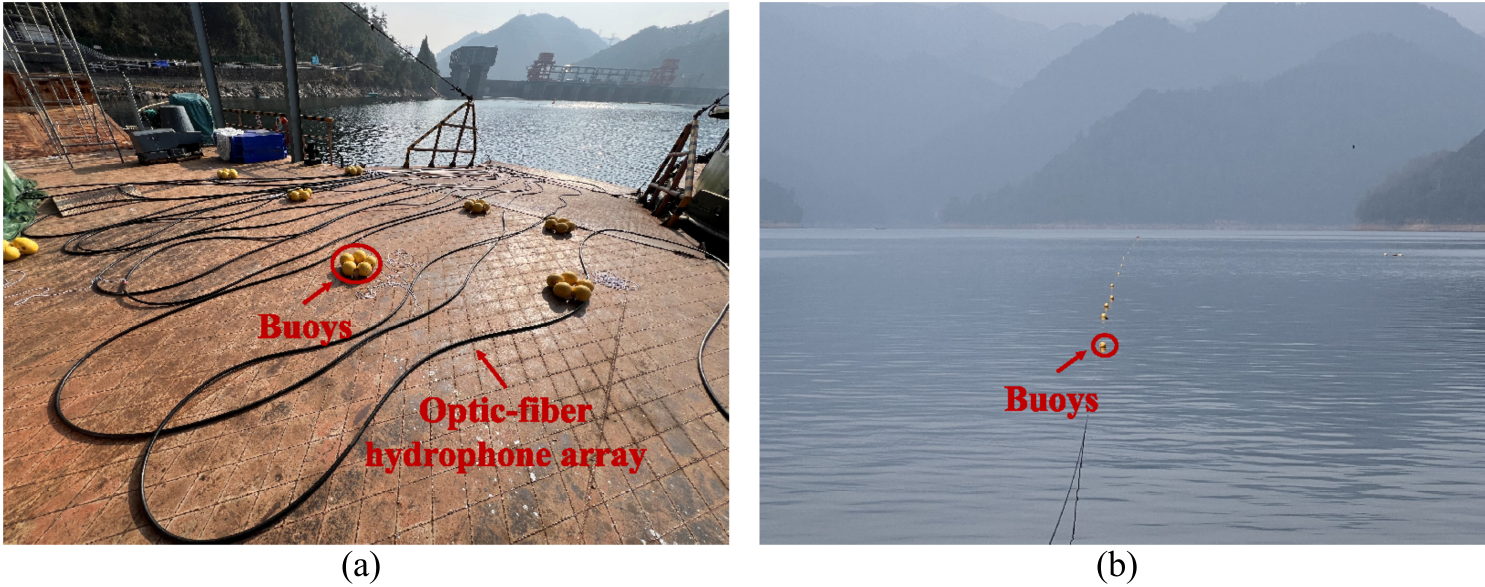}\label{fig: deployment scene of array}
\caption{Deployment of the fiber-optic hydrophone array in the lake trial: (a) Buoys and fiber-optic array during deployment. (b) Scene after the array deployment.}
\label{fig: deployment of array}
\vspace{-0.4cm}
\end{figure}

Second, we assess the efficacy of the long array segmentation strategy on a $100$ elements array. As introduced in Sec. \ref{sec: array shape calibration method}, a sliding-window segmentation paradigm with overlap elements is employed to alleviate the dimensionality burden associated with large scale arrays. The choice of overlap ratio presents a balance between the global alignment accuracy and the computational expense. To this end, we analyze both the geometric RMSE and DoA MAE against varying degrees of sub-array overlap from $10\%$ to $50\%$. For each ratio, $N=50$ independent Monte Carlo trials are executed.

In Fig. \ref{fig: MC results overlap}, we demonstrate the error metrics at different overlap ratios. Given low overlap ratio values, especially at $10\%$ and $20\%$, the shared geometric constraints between adjacent sub-arrays are insufficient. The weak coupling fails to suppress the accumulation of estimation errors during the global alignment phase, resulting in severe geometric mismatch and severe DoA MAE exceeding $13^\circ$. An inflection point emerges at an overlap ratio of $30\%$, where the geometric RMSE decreases to $0.2$ meters and the DoA MAE converges to nearly $0^\circ$. Further increasing the overlap produce only small precision gains but require more repeated processing. Consequently, an overlap ratio of $30\%$ provides a favorable trade-off between estimation accuracy and computational cost.

\section{Real-World Validation}\label{sec: real-world validations}
In this section, we evaluate BOGE with two sets of data collected in field experiments. The public dataset SWellEx-96 provides reference element coordinates for the evaluation of geometric reconstruction. A lake trial using an self-made array further evaluates the performance of fixed source localization and moving source tracking. Note that ASMLM assumes the array geometry can be represented by a single circular-arc parameter, and WORKS relies on multiple reliable spectral lines. These assumptions are not fully satisfied in both SWellEx-96 dataset and our lake trial setting. Therefore, we compare BOGE with AS-TDE and DAS-SC in real-world validation.

\subsection{Validation on the SWellEx-96 Dataset}\label{sec: swellex96 validation}
The SWellEx-96 experiment was conducted from May 10 to 18, 1996, approximately $12$ km off Point Loma near San Diego, California. We applied BOGE to Event S5 of the SWellEx-96 dataset and report the result obtained from the North HLA data at $166$ Hz. This HLA consists of $27$ effective elements, which were deployed at approximately the same depth, so we performed the calibration in 2D space. To account for nonuniform spacing and invalid elements, BOGE used the individual link lengths and the coordinates of the first, 14th, and 27th effective elements. The coordinates of the remaining elements were excluded from calibration and used only for evaluation after calibration.

\begin{figure}[t]
\centering
\includegraphics[width=\columnwidth]{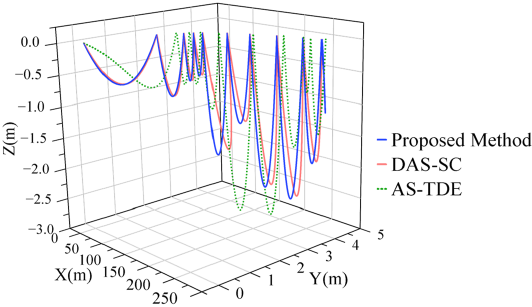}
\caption{Reconstructed Shape comparison between BOGE, DAS-SC, and AS-TDE.}
\label{fig: lake trial shape comparison}
\vspace{-0.4cm}
\end{figure}

Fig. \ref{fig: swellex96 result}(a) compares the geometry consistency performance of all schemes, which shows that BOGE aligns better with the reference shape than other benchmarks. Further validated in Fig. \ref{fig: swellex96 result}(b), BOGE achieves a geometric RMSE of $0.659$ meters for the reconstruction, compared with $4.095$ m for DAS-SC and $5.584$ m for AS-TDE. The coordinates RMSE between the North HLA estimates obtained at $166$ and $201$ Hz is $0.580$ meters. This consistency across frequencies supports the stability of the reconstruction under the tested conditions.

\subsection{Lake Trial Setup}\label{Lake Trial setup}
The experiments were carried out at a lake in Hangzhou, China, in January 2024. As shown in Fig. \ref{fig: schematic diagram of array layout}, the testing array was a flexible fiber-optic hydrophone array consisting of $M=238$ elements, with a uniform nominal element-spacing of $d=1$ meter. The array was suspended at a nominal depth of $15$ meters below the water surface via ten buoys. The buoys were deployed with uneven horizontal spacing, attached to the array at specific element indices: 0, 20, 40, 50, 60, 90, 120, 160, 200, and 230. The coordinates of the ten elements were known and set as reference elements, providing stronger constraints. Uneven spacing between the buoys allowed the section to sag under gravity and buoyancy. The global geometry of the testing array thereby formed a multi-section catenary-like 3D deformation along the $240$ meters. The deployment process of the array is illustrated in Fig. \ref{fig: deployment of array}. The benchmark geometry was reconstructed using the AS-TDE described by \cite{jun2007method}.

The acoustic sources were positioned in the far-field and transmitted signals from different directions, including $-30^\circ$ and $-5^\circ$. The analog signals received by the hydrophones were digitized by a multi-channel data acquisition system with a sampling rate of $10$ kHz.

\begin{figure}[t]
\centering
\includegraphics[width=\columnwidth]{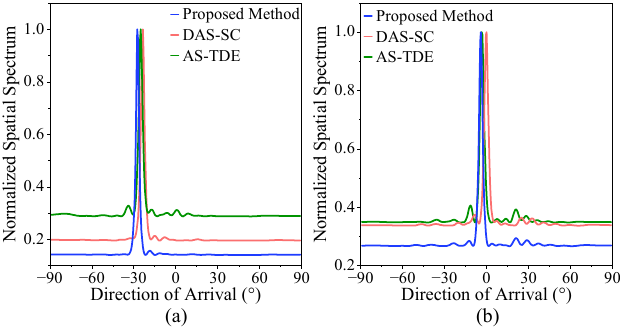}\label{fig: fixed source doa}
\caption{Normalized MUSIC spectra for fixed targets in different directions: (a) Source direction from $-30^\circ$. (b) Source direction from $-5^\circ$.}
\label{fig: lake-trial doa for a fixed target}
\vspace{-0.4cm}
\end{figure}

\subsection{Array Shape Calibration}\label{sec: array shape calibration}
To validate the practical performance of BOGE, the lake-trial data was processed and analyzed. Fig. \ref{fig: lake trial shape comparison} illustrates the reconstructed spatial geometry of the hydrophone array. The shape derived from the AS-TDE and DAS-SC calibration methods are shown by the green line and the red line, while the geometry estimated by the proposed method is represented by the blue line. The proposed framework reconstructs the multi-section catenary shape induced by ten suspension buoys, which is consistent with the deployment schematic in Fig. \ref{fig: schematic diagram of array layout}. The estimated geometry follows the overall array profile and captures the main localized sagging sections. Since an absolute ground-truth shape for such a long array is unavailable in a lake environment, an evaluation of the calibration accuracy necessitates a comparison of the subsequent DoA estimation performance.

\subsection{DoA Estimation for a Fixed Target}\label{doa estimation fof fixed targets}
The narrow-band MUSIC algorithm is applied to locate a fixed acoustic source positioned at different directions, with transmission frequencies of $500$ Hz.

Fig. \ref{fig: lake-trial doa for a fixed target} shows the normalized MUSIC spectra. The spectra established by the AS-TDE and the DAS-SC are denoted by the green lines and red lines, and the results achieved by BOGE are represented by the blue lines.

All three methods obtain identifiable main lobes at both source directions. Specifically, the spatial spectra generated by BOGE demonstrate more concentrated peaks and lower side-lobe levels under the tested conditions. These results indicate that BOGE can provide reliable DoA estimation in complex real-world underwater environments.

\subsection{DoA Tracking for a Moving Target}\label{sec: doa tracking for a moving target}

Building upon the successful localization of static sources, the dynamic tracking capability of the calibrated array is further evaluated. A moving acoustic source emitting at $500$ Hz was deployed in the lake trial and its spatial trajectory was continuously monitored during a 60-second observation window.

\begin{figure}[t]
\centering
\includegraphics[width=\columnwidth]{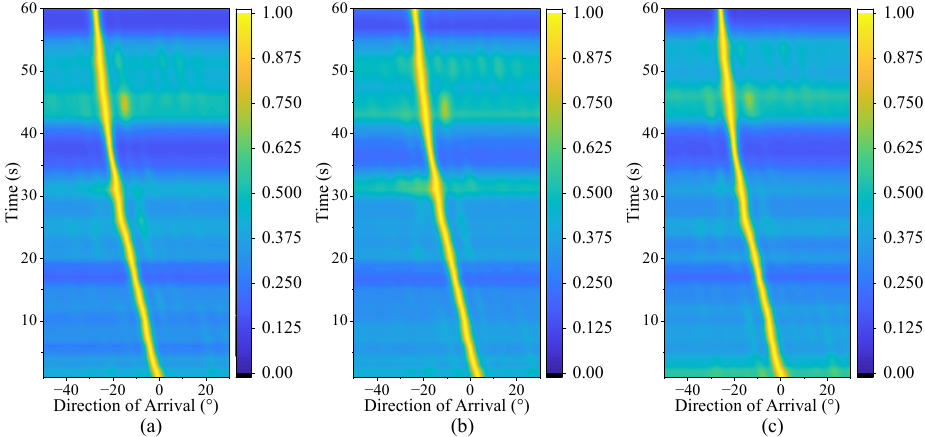}\label{fig: doa for a moving target}

\caption{Performance on tracking a moving target: (a) Tracking trajectory using BOGE. (b) Tracking trajectory using DAS-SC. (c) Tracking trajectory using AS-TDE.}
\label{fig: lake-trial doa for a moving target}
\end{figure}

As shown in Fig. \ref{fig: lake-trial doa for a moving target}, the DoA tracking results are visualized as normalized bearing-time records (BTR). Specifically, Fig. \ref{fig: lake-trial doa for a moving target}(a) demonstrates the tracking trajectory established by BOGE, while Fig. \ref{fig: lake-trial doa for a moving target}(b) and Fig. \ref{fig: lake-trial doa for a moving target}(c) present the result obtained by DAS-SC and AS-TDE. 

As shown in the BTR plots, the target exhibits a continuous spatial transition from $0^\circ$ to approximately $-30^\circ$ over 60 seconds duration. All methods successfully capture the dynamic movement. Comparing the subplots, the tracking result generated by BOGE illustrates a more focused main-lobe trajectory. Table \ref{tab:comparison} provides a quantitative comparison over the same interval. BOGE yields a main lobe width of $4.87^\circ$, compared with $8.98^\circ$ for DAS-SC and $13.91^\circ$ for AS-TDE. The corresponding noise levels are similar at $-4.91$, $-4.50$ and $-4.77$ dB, respectively. The dynamic tracking performance proves that the proposed method provides a stable and reliable constructed geometry for continuous acoustic monitoring under the tested conditions, which is comparable to the comparison methods.

\section{Conclusion}
In this paper, we proposed BOGE, a self-calibration framework that combines a physics-informed parametric model with a BO-guided hierarchical optimization strategy. Overlapping sub-array alignment further extends the framework to long arrays. In the SNR benchmark from $-20$ to $20$ dB, with 200 trials at each SNR, BOGE achieved the lowest mean geometric RMSE over most of the tested range. On the public SWellEx-96 dataset, BOGE achieved a geometric RMSE of $0.659$ m at $166$ Hz, while the coordinate RMSE between the estimates at $166$ and $201$ Hz was $0.580$ m. A lake trial with a fiber-optic hydrophone array consisting of 238 elements yielded fixed-source localization and moving-target tracking performance comparable to the comparison methods. For the moving source, BOGE achieves the narrowest main lobe. These results support BOGE as a practical calibration approach without cooperative sources.

\begin{table}[t]
\centering
\caption{Comparison of moving source tracking performance.}
\label{tab:comparison}
\renewcommand{\arraystretch}{1.2}
\setlength{\tabcolsep}{12pt}
\begin{tabular}{cccc}
\toprule
Metric & BOGE & DAS-SC & AS-TDE \\
\midrule
main-lobe width ($^\circ$)
    & $4.87$
    & $8.98$
    & $13.91$ \\
noise level (dB)
    & $-4.91$
    & $-4.50$
    & $-4.77$ \\
\bottomrule
\end{tabular}
\end{table}

Future work will extend BOGE to online calibration of towed arrays, which requires BO and MUSIC evaluations to follow geometry changes simultaneously within a short time. This line of works is expected to model continuous 3D deformation caused by vessel maneuvers and currents \cite{lan2024array}. Another challenge is jointly tracking time-varying array geometry and source DoAs under limited snapshots without imposing restrictive deformation assumptions \cite{wang2024fast,wang2025adaptive}.


\bibliographystyle{ieeetr}
\bibliography{refer}

@article{miao2026shape,
  title={Shape self-calibration method of the distributed acoustic sensing cable for beamforming enhancement},
  author={Miao, Qiuyan and Yan, Guofeng and Jiang, Lang and Xiao, Chun and Guo, Ruqian and Hu, Weiwang and Zhao, Chunliu},
  journal={Optics \& Laser Technology},
  volume={193},
  pages={114243},
  year={2026}
}

@article{sheikh2016near,
  title={Near field source localization in the presence of array sensor position uncertainties},
  author={Sheikh, Yawar Ali and Ye, Zhongfu and Ullah, Rizwan and Shabir, Kashif and Luo, Dawei},
  journal={Communications on Applied Electronics},
  volume={6},
  number={3},
  pages={1--6},
  year={2016}
}

@article{wang2013tdoa,
  title={{TDOA} source localization in the presence of synchronization clock bias and sensor position errors},
  author={Wang, Yue and Ho, KC},
  journal={IEEE Transactions on Signal Processing},
  volume={61},
  number={18},
  pages={4532--4544},
  year={2013}
}

@article{pan2024fast,
  title={Fast estimation of array shape and direction of arrival using Sparse {B}ayesian Learning for manoeuvring towed line array},
  author={Pan, Xiang and Wang, Haoran and Li, Min and Zhou, Jie and Li, Yuxiao and Xu, Weize},
  journal={IET Radar, Sonar \& Navigation},
  volume={18},
  number={10},
  pages={1625--1637},
  year={2024}
}

@article{chen2024distributed,
  title={Distributed shape detection for an acoustic sensitive optical cable with {DAS}},
  author={Chen, Boqi and Wang, Zhaoyong and Yang, Junqi and Liu, Yifan and Chen, Yici and Wu, Jinyi and Gao, Kan and Ye, Qing},
  journal={Optics Letters},
  volume={49},
  number={12},
  pages={3384--3387},
  year={2024}
}

@article{wang2023self,
  title={Self-calibration method of sensors array errors based on rotation measurement},
  author={Wang, Ke and Cheng, Feng and Yi, Jianxin and Wan, Xianrong},
  journal={IEEE Sensors Journal},
  volume={23},
  number={3},
  pages={2311--2319},
  year={2023}
}

@article{wan2014identifiability,
  title={Identifiability analysis for array shape self-calibration based on hybrid {Cram{\'e}r-Rao} bound},
  author={Wan, Shuang and Tang, Jun and Zhu, Wei and Zhang, Ning},
  journal={IEEE Signal Processing Letters},
  volume={21},
  number={4},
  pages={473--477},
  year={2014}
}

@article{yang2019joint,
  title={Joint calibration of array shape and sensor gain/phase for highly deformed arrays using wideband signals},
  author={Yang, Long and Yang, Yixin and Liao, Guisheng and Guo, Xijing},
  journal={Signal Processing},
  volume={165},
  pages={222--232},
  year={2019}
}

@article{song2024sensor,
  title={Sensor position self-calibration for nominal linear array under small positional error},
  author={Song, Shuoshuo and Ma, Xiaofeng and Zhou, Siyi and Sheng, Weixing},
  journal={IEEE Transactions on Aerospace and Electronic Systems},
  volume={60},
  number={5},
  pages={7484--7490},
  year={2024}
}

@article{zhang2024array,
  title={Array shape calibration based on coherence of noise radiated by non-cooperative ships},
  author={Zhang, Wenchang and Jiang, Pengfei and Lin, Jianheng and Sun, Junping},
  journal={Ocean Engineering},
  volume={303},
  pages={117792},
  year={2024}
}

@article{gharib2024modeling,
  title={Modeling and analysis of static and dynamic behavior of marine towed cable-array system based on the vessel motion},
  author={Gharib, Mohammad Reza and Heydari, Ali and Salehi Kolahi, Mohammad Reza},
  journal={Advances in Mechanical Engineering},
  volume={16},
  number={1},
  pages={16878132231220353},
  year={2024}
}

@article{odom2014passive,
  title={Passive towed array shape estimation using heading and acoustic data},
  author={Odom, Jonathan L and Krolik, Jeffrey L},
  journal={IEEE Journal of Oceanic Engineering},
  volume={40},
  number={2},
  pages={465--474},
  year={2015}
}

@article{wu2021enhanced,
  title={An enhanced data-driven array shape estimation method using passive underwater acoustic data},
  author={Wu, Qisong and Zhang, Hao and Lai, Zhichao and Xu, Youhai and Yao, Shuai and Tao, Jun},
  journal={Remote Sensing},
  volume={13},
  number={9},
  pages={1773},
  year={2021}
}

@article{friedlander2002sensitivity,
  title={A sensitivity analysis of the {MUSIC} algorithm},
  author={Friedlander, Benjamin},
  journal={IEEE Transactions on Acoustics, Speech, and Signal Processing},
  volume={38},
  number={10},
  pages={1740--1751},
  year={1990}
}

@article{rockah2003array,
  title={Array shape calibration using sources in unknown locations--Part I: Far-field sources},
  author={Rockah, Yosef and Schultheiss, P},
  journal={IEEE Transactions on Acoustics, Speech, and Signal Processing},
  volume={35},
  number={3},
  pages={286--299},
  year={1987}
}

@inproceedings{santori2007array,
  title={Array shape self-calibration for large flexible antenna},
  author={Santori, Agnes and Barrere, Jean and Chabriel, Gilles and Jauffret, Claude and Medynski, Dominique},
  booktitle={2007 {IEEE} Aerospace Conference},
  pages={1--9},
  year={2007}
}

@article{li2006theoretical,
  title={Theoretical analyses of gain and phase error calibration with optimal implementation for linear equispaced array},
  author={Li, Youming and Er, Meng Hwa},
  journal={IEEE Transactions on Signal Processing},
  volume={54},
  number={2},
  pages={712--723},
  year={2006}
}

@book{williams2006gaussian,
  title={Gaussian processes for machine learning},
  author={Williams, Christopher K. I. and Rasmussen, Carl Edward},
  year={2006},
  publisher={MIT Press},
  address={Cambridge, MA}
}

@article{friedlander2002direction,
  title={Direction finding in the presence of mutual coupling},
  author={Friedlander, Benjamin and Weiss, Anthony J},
  journal={IEEE Transactions on Antennas and Propagation},
  volume={39},
  number={3},
  pages={273--284},
  year={1991}
}

@article{weiss2002array,
  title={Array shape calibration using sources in unknown locations-a maximum likelihood approach},
  author={Weiss, Anthony J and Friedlander, Benjamin},
  journal={IEEE Transactions on Acoustics, Speech, and Signal Processing},
  volume={37},
  number={12},
  pages={1958--1966},
  year={1988}
}

@article{mohsan2023recent,
  title={Recent advances, future trends, applications and challenges of {Internet of Underwater Things} ({IoUT}): A comprehensive review},
  author={Mohsan, Syed Agha Hassnain and Li, Yanlong and Sadiq, Muhammad and Liang, Junwei and Khan, Muhammad Asghar},
  journal={Journal of Marine Science and Engineering},
  volume={11},
  number={1},
  pages={124},
  year={2023}
}

@article{zheng2020joint,
  title={Joint towed array shape and direction of arrivals estimation using sparse {B}ayesian learning during maneuvering},
  author={Zheng, Zheng and Yang, TC and Gerstoft, Peter and Pan, Xiang},
  journal={The Journal of the Acoustical Society of America},
  volume={147},
  number={3},
  pages={1738--1751},
  year={2020}
}

@article{ramamohan2022self,
  title={Self-calibration of acoustic scalar and vector sensor arrays},
  author={Ramamohan, Krishnaprasad Nambur and Chepuri, Sundeep Prabhakar and Comesa{\~n}a, Daniel Fernandez and Leus, Geert},
  journal={IEEE Transactions on Signal Processing},
  volume={71},
  pages={61--75},
  year={2022}
}

@article{liu2016sparse,
  title={A sparse-based approach for {DOA} estimation and array calibration in uniform linear array},
  author={Liu, Hongqing and Zhao, Luming and Li, Yong and Jing, Xiaorong and Truong, Trieu-Kien},
  journal={IEEE Sensors Journal},
  volume={16},
  number={15},
  pages={6018--6027},
  year={2016}
}

@article{xenaki2025overview,
  title={Overview of distributed acoustic sensing: Theory and ocean applications},
  author={Xenaki, Angeliki and Gerstoft, Peter and Williams, Ethan and Abadi, Shima},
  journal={The Journal of the Acoustical Society of America},
  volume={158},
  number={1},
  pages={801--825},
  year={2025}
}

@article{jahanbakht2021internet,
  title={Internet of underwater things and big marine data analytics—a comprehensive survey},
  author={Jahanbakht, Mohammad and Xiang, Wei and Hanzo, Lajos and Azghadi, Mostafa Rahimi},
  journal={IEEE Communications Surveys \& Tutorials},
  volume={23},
  number={2},
  pages={904--956},
  year={2021}
}

@article{li2025underwater,
  title={Underwater acoustic communications},
  author={Li, Zhengnan and Chitre, Mandar and Stojanovic, Milica},
  journal={Nature Reviews Electrical Engineering},
  volume={2},
  number={2},
  pages={83--95},
  year={2025}
}

@article{zhang2022robust,
  title={Robust underwater direction-of-arrival tracking with uncertain environmental disturbances using a uniform circular hydrophone array},
  author={Zhang, Boxuan and Hou, Xianghao and Yang, Yixin},
  journal={The Journal of the Acoustical Society of America},
  volume={151},
  number={6},
  pages={4101--4113},
  year={2022}
}

@inproceedings{zheng2019towed,
  title={Towed array beamforming using sparse {B}ayesian learning during maneuvering},
  author={Zheng, Zheng and Yang, TC and Pan, Xiang and Gerstoft, Peter},
  booktitle={{OCEANS} 2019-Marseille},
  pages={1--6},
  year={2019}
}

@inproceedings{santori2009sensor,
  title={Sensor self-calibration methods for a passive conformal airborne antenna},
  author={Santori, Agn{\`e}s},
  booktitle={2009 International Radar Conference" Surveillance for a Safer World"(RADAR 2009)},
  pages={1--5},
  year={2009}
}

@article{liu2018wideband,
  title={Wideband array self-calibration and {DOA} estimation under large position errors},
  author={Liu, Yaqi and Liu, Chengcheng and Zhao, Yongjun and Zhu, Jiandong},
  journal={Digital Signal Processing},
  volume={78},
  pages={250--258},
  year={2018},
}

@article{zhen2018array,
  title={Array calibration method in super-resolution direction finding for wideband signals},
  author={Zhen, Jiaqi and Li, Yanchao},
  journal={Journal of Information Hiding and Multimedia Signal Processing},
  volume={9},
  number={1},
  pages={129--144},
  year={2018}
}

@article{schmidt1986multiple,
  title={Multiple emitter location and signal parameter estimation},
  author={Schmidt, Ralph},
  journal={IEEE Transactions on Antennas and Propagation},
  volume={34},
  number={3},
  pages={276--280},
  year={1986}
}

@article{akyildiz2005underwater,
  title={Underwater acoustic sensor networks: Research challenges},
  author={Akyildiz, Ian F and Pompili, Dario and Melodia, Tommaso},
  journal={Ad Hoc Networks},
  volume={3},
  number={3},
  pages={257--279},
  year={2005}
}

@article{heidemann2012underwater,
  title={Underwater sensor networks: Applications, advances and challenges},
  author={Heidemann, John and Stojanovic, Milica and Zorzi, Michele},
  journal={Philosophical Transactions of the Royal Society A: Mathematical, Physical and Engineering Sciences},
  volume={370},
  number={1958},
  pages={158--175},
  year={2012}
}

@article{fischer2020operating,
  title={Operating cabled underwater observatories in rough shelf-sea environments: A technological challenge},
  author={Fischer, Philipp and Brix, Holger and Baschek, Burkard and Kraberg, Alexandra and Brand, Markus and Cisewski, Boris and Riethm{\"u}ller, Rolf and Breitbach, Gisbert and M{\"o}ller, Klas Ove and Gattuso, Jean-Pierre and others},
  journal={Frontiers in Marine Science},
  volume={7},
  pages={551},
  year={2020}
}

@article{wang2024fast,
  title={Fast joint estimation of direction of arrival and towed array shape based on marginal likelihood maximization},
  author={Wang, Junxiong and Pan, Xiang and Li, Ao and Liu, Fenting and Jiao, Jianbo},
  journal={Digital Signal Processing},
  volume={154},
  pages={104676},
  year={2024}
}

@article{lan2024array,
  title={Array shape estimation based on tug vehicle noise for towed linear array sonar during turning},
  author={Lan, Tian and Wang, Yilin and Qiu, Longhao and Liu, Guolong},
  journal={Ocean Engineering},
  volume={303},
  pages={117554},
  year={2024}
}

@article{wang2025adaptive,
  title={Adaptive array shape estimation and high-resolution sensing for {AUV}-towed linear array sonar during turns},
  author={Wang, Junxiong and Pan, Xiang and Cheng, Lei and Jiao, Jianbo},
  journal={Remote Sensing},
  volume={17},
  number={15},
  pages={2690},
  year={2025}
}

@article{jun2007method,
  title={A method of array shape calibration based on time delay estimation using two auxiliary sources},
  author={Wang, Jun and Wu, Lixin and Lynch, Jim and Newhall, Arthur},
  journal={Acta Acustica},
  volume={32},
  number={2},
  pages={165--170},
  year={2007}
}

@article{yang2025shape,
  title={Shape self-calibration of a highly deformed sonar array by adding heading sensors},
  author={Yang, Yixin and Zhou, Bangjie and Yang, Long and Yu, Mengling},
  journal={Measurement},
  pages={118988},
  year={2025}
}

@article{zhang2026multi,
  title={Multi-fidelity {B}ayesian optimization for {N}ash equilibria with black-box utilities},
  author={Zhang, Yunchuan and Simeone, Osvaldo and Poor, H Vincent},
  journal={IEEE Transactions on Signal Processing},
  volume={74},
  pages={1015-1029},
  year={2026}
}

\newpage

 





\end{document}